\documentclass{iopart}
\usepackage{iopams}
\expandafter\let\csname equation*\endcsname\relax
\expandafter\let\csname endequation*\endcsname\relax
\usepackage{amsmath}
\usepackage{graphicx}

\usepackage{color}
 
\newcommand{\Mi}{M_\mathrm{in}}
\newcommand{\Mp}{M_\mathrm{ph}}

\begin{document}
\title{Stability and dispersion relations of three-dimensional solitary waves
in trapped Bose-Einstein condensates}
\author{A. Mu\~{n}oz Mateo}
\ead{amunoz@ecm.ub.edu}
\address{Departament d'Estructura i Constituents de la Mat\`{e}ria,
Facultat de F\'{i}sica, Universitat de Barcelona, E–08028 Barcelona, Spain}
\address{Centre for Theoretical Chemistry and Physics and New Zealand 
Institute for Advanced Study, Massey University, Private Bag 102904 NSMC, 
Auckland 0745, New Zealand}

\author{J. Brand}
\ead{J.Brand@massey.ac.nz}
\address{Dodd-Walls Centre for Photonic and Quantum 
Technologies and Centre for Theoretical Chemistry and Physics, New Zealand 
Institute for Advanced Study, Massey University, Private Bag 102904 NSMC, Auckland 0745, New 
Zealand}


\begin{abstract}
We analyse the dynamical properties of three-dimensional solitary waves in
cylindrically trapped Bose-Einstein condensates. Families of solitary waves 
bifurcate from the planar dark soliton and include the solitonic vortex, the 
vortex ring and more complex structures of intersecting vortex-line known collectively as Chladni solitons.
The particle-like dynamics of these  guided solitary waves provides potentially
profitable features for their implementation in atomtronic circuits, and play
a key role in the generation of metastable loop currents.
Based on the time-dependent Gross-Pitaevskii equation we calculate the 
dispersion relations of moving solitary waves and their modes of dynamical 
instability. The dispersion relations reveal a complex crossing and bifurcation 
scenario. For  stationary structures we find that for 
$\mu/\hbar\omega_\perp > 2.65$ the solitonic vortex is the only stable solitary 
wave. More complex Chladni solitons still have weaker instabilities than planar 
dark solitons and may be seen as transient structures in experiments. Fully 
time-dependent simulations illustrate typical decay scenarios, which may result 
in the generation of multiple separated solitonic vortices.


\end{abstract}
\maketitle

\section{Introduction}

Nonlinear waves in superfluids are the subject of intense theoretical and 
experimental research. 
The exquisite
control achieved in manipulating ultracold atomic gases has enabled the 
creation, manipulation, and detection of dark solitons \cite{Burger1999,Denschlag2000}, vortex rings 
\cite{Anderson2001,ginsberg05} and solitonic vortices in Bose-Einstein condensates 
(BECs) \cite{Donadello2014} 
and Fermi gases along the BEC--BCS crossover \cite{Ku2014}. All these structures are strongly influenced by the confinement of the particle cloud and represent solitary waves in the sense that they are characterised by an excitation energy density above the condensate ground state that is localised with respect to the long axis of the confining geometry. These multi-dimensional solitary waves distinguish themselves by their non-trivial topology associated with the constituting superfluid currents.


The combination of confinement and superfluid currents is also the main
constituent in the development of atomtronic devices, and, to this end, an in-depth understanding
of the nonlinear phenomena involved in such dynamics is required.
In particular, the role played by nonlinear waves deserves special attention.
On the one hand, the shape-preserving evolution of solitary waves, in both 
repulsively and attractively interacting systems, could be a useful
feature to be implemented in future applications, in a similar way as 
optical solitons are being currently used in optical fibers \cite{Mollenauer2006}.
On the other hand, the necessity to better  understand  the role
played by solitary waves in the generation of superfluid currents has manifested itself 
in a series of experiments with superfluid rings at 
NIST \cite{Wright2013,Eckel2014}. Therein,
vortices of solitonic nature, due to the transverse trapping along the radius of the rings,
have been found after driving the superfluid into motion. Additionally, the 
generation of such metastable loop currents has been demonstrated to be mediated by
the existence of solitary waves that produce an energy barrier preventing
phase slips \cite{Mateo2015}.

Dark solitons, or 
\emph{kinks}, are density dips with an associated jump in the phase of the 
order 
parameter, and represent nonlinear excitations in Bose-Einstein condensates with repulsive 
interparticle interactions \cite{Frantzeskakis2010}.
In one dimensional rings, kink excitations represent 
intermediate stages connecting states with different winding numbers 
\cite{Kanamoto2008}. A 
one dimensional dark soliton can be understood as a vortex which is crossing the 
ring, and hence providing a characteristic density depletion and phase slip 
that depends on the position of the vortex. In higher dimensions, the 
structure of a vortex line crossing the ring, a solitonic vortex, can be more 
easily identified on the cross section of the system. Alternatively, other 
transverse states containing vortex lines can be excited in order to produce a 
given phase jump along the ring circumference. In general, multidimensional 
stationary kinks are dynamically unstable \cite{Muryshev1999}, unless a tight 
trap could 
keep the system in the quasi-one dimensional regime so that higher energy 
transverse excitations were excluded \cite{brand01a,Brand2002}.

In trapped superfluids, 
Chladni solitons \cite{Mateo2014} emerge from the decay of three dimensional 
kinks, 
as a result of the excitation of standing waves on the nodal plane of the kink. 
Such waves produce patterns of vorticity along the nodal lines of the 
transverse modes in analogy to the Chladni figures visualising the nodal lines of plate vibration modes \cite{Chladni1787}.
In traps with cylindrical symmetry, the different families of solitary waves 
can be described by the radial $p$ and azimuthal $l$ quantum numbers, 
indicating the number of transverse nodal lines along their respective 
directions. Solitonic vortices, belonging to the family $(p=0,l=1)$, are 
the lowest energy states in elongated condensates \cite{Brand2002,Komineas2003}, 
while vortex rings \cite{Komineas2002,Komineas2003a}, presenting higher excitation energies, belong to the 
family $(p=1,l=0)$. Along with these previously known states, there exist a sequence of more complex
stationary solitary waves with all possible combinations of $p$ and $l$ quantum numbers, as was  pointed out by the authors \cite{Mateo2014}. Very recently evidence for the observation of the $\Phi$-shaped Chadni soliton with quantum numbers $(p=1,l=1)$  in a superfluid Fermi gas at unitary was reported in Ref.~\cite{Ku}.
The relative strength of the decay modes that can produce Chladni solitons from the decay of the kink, as well as the robustness of the Chladni solitons are key points that remain to be clarified.

Motivated by the previous considerations, in the present work we study the
dynamics of solitary wave excitations within 
the framework of the time-dependent Gross-Pitaevskii equation. Section \ref{sec:mass} is devoted to characteristic mass parameters relevant for the Landau quasiparticle dynamics of solitary waves. Numerical data is presented and compared to analytical approximations for energy, inertial, and physical masses of the dark soliton in subsection \ref{sec:3Dkink}, and the solitonic vortex and vortex ring in subsection \ref{sec:SV}. The snaking instability of the kink state is analysed in detail in section \ref{sec:SI} by solving the Bogoliubov equations of linearised excitations for the trapped kink state numerically and by developing a semi-analytical theory of the unstable modes. Numerical results for stationary Chladni solitons are reported in section \ref{sec:Chladni}, while the dispersion relations and phase step of moving Chladni solitons are considered in section \ref{sec:Moving}. A stability analysis of Chladni solitons -- stationary and moving -- is performed in section \ref{sec:Stability}, where also results from real-time evolution beyond the linear response regime are reported.
%
We identify two 
characteristic scenarios in the fate of Chladni solitons: either a chain of 
decay episodes into single vortex lines which are localized around a transverse plane of the 
system, or the generation of secondary travelling waves. 
Finally, the dynamical features pointed out in our study are 
used to propose feasible protocols for the experimental realization of these 
solitary waves.  

\section{Energy and inertial and physical masses} \label{sec:mass}

Solitary waves often exhibit particle-like dynamics. One manifestation of such particle-like dynamics occurs when solitary waves move across a slowly varying background where energy radiation is being suppressed. As a consequence of energy conservation, the solitary wave will then adapt adiabatically to the changing environmental conditions, adjusting its internal parameters as to maintain its local energy constant, and acting as a Landau quasiparticle \cite{Konotop2004}. The nonlinear wave solutions considered in this article are all solitary waves in this sense, because they are localised with respect to the long axis of a trapped geometry and thus may perform guided quasiparticle motion, even though their properties are significantly influenced by the presence of a transverse confining potential. In sections \ref{sec:SI} and \ref{sec:Stability} we present evidence to show that only two types of solitary waves are dynamically stable in cylindrically trapped Bose-Einstein condensates: The solitonic vortex is stable when $\mu/(\hbar\omega_\perp) > 2.65$ while the dark soliton is stable below this value, where $\mu$ is the chemical potential and $\omega_\perp$ the frequency of the transverse harmonic trapping potential. Dynamically stable solitary wave can be expected to perform near-hamiltonian quasiparticle dynamics for a long time and have been observed in experiments with trapped superfluid Fermi gases for several seconds \cite{Yefsah,Ku2014}. Unstable solitary waves like vortex rings may still exhibit quasiparticle like dynamics if the competing decay dynamics is slow enough or suppressed by symmetry constraints \cite{Reichl2013,Bulgac2013}.

In the framework of Landau quasiparticle dynamics, the equations of motion of a solitary wave in a trapped quantum gas can be derived from knowing the excitation energy $E_s(\mu,v_s)$ of the solitary wave as a function of the chemical potential $\mu$ and its velocity $v_s$ \cite{Konotop2004}. In a trapped gas, the chemical potential is then treated as a (slowly varying) function of the position $Z$ of the solitary wave, while the velocity is the time derivative of position $v_s = \dot{Z}$. Requiring the energy to be a constant of motion then leads to 
\begin{align} \label{eq:eom1}
0= \frac{dE_s}{dt} = \left(\left.\frac{\partial E_s}{\partial \mu}\right|_{v_s} \frac{d \mu}{dZ} + \Mi \ddot{Z}\right)\dot{Z} ,
\end{align}
where 
\begin{align}
\Mi = \frac{1}{v_s} \left.\frac{\partial E_s}{\partial v_s}\right|_{\mu} ,
\end{align}
is the \emph{inertial mass} of the solitary wave, often also called the \emph{effective mass}. Equation \eqref{eq:eom1} already looks like Newton's law. For  weak harmonic trapping potential along the $z$ axis where the Thomas Fermi approximation demands that $\mu(Z) = \mu_0 - \frac{1}{2}m \omega_z^2 Z^2$, we arrive at Newton's equation of motion in the form
\begin{align} \label{eq:eom2}
 \Mi \ddot{Z} = - \Mp \omega_z^2 Z ,
\end{align}
where  the \emph{physical mass} defined by
\begin{align}
\Mp = m \left.\frac{\partial E_s}{\partial \mu}\right|_{v_s} ,
\end{align}
is to be interpreted characteristic parameter of the solitary wave that gives rise to the buoyancy-like force on the right hand side of Eq.~\eqref{eq:eom2}. The interpretation of the physical mass to be related to a buoyancy phenomenon is further supported by it being closely related to number of missing particles $N_s$, i.e.\ particles depleted from the background density due to the presence of the solitary wave, where $\Mp=m N_s$ holds in many cases \cite{Scott2011}.

Solitary waves in repulsively interacting quantum gases typically have negative inertial and physical masses, which leads to oscillatory motion of the solitary waves in a trapped gas. This, e.g.\ is the case for the one-dimensional Gross-Pitaevskii equation describing Bose-Einstein condensates with tight transverse confinement \cite{Busch2000,Konotop2004}. If the physical and inertial masses are independent of position and velocity, or in the limit of small-amplitude motion, we obtain simple harmonic oscillations $Z(t)\propto \sin(\Omega t)$ with \cite{Scott2011,Liao11pr:FermiSolitons}
\begin{align}
\frac{\omega_z^2}{\Omega^2} = \frac{\Mi}{\Mp} .
\end{align}
Such harmonic oscillations have already been observed in experiments and the frequency ratio measured for dark solitons \cite{Becker2008} and for solitonic vortices \cite{Serafini2015}  in Bose-Einstein condensates and for solitonic vortices in the superfluid Fermi gas \cite{Yefsah,Ku2014}. 
In the remainder of this section we present numerical data of dispersion relations and mass parameters evaluated for the dark soliton, solitonic vortex and vortex rings, and compare with approximate analytical expressions.

\subsection{Planar dark solitons}\label{sec:3Dkink}

We will model solitary waves within the mean field theory given by the 
Gross-Pitaevskii equation for the condensate order parameter 
$\Psi(\mathbf{r},t)$
\begin{equation}
i\hbar\frac{\partial\Psi}{\partial t}=\left(  -\frac{\hbar^{2}}{2m}\nabla
^{2}+V(\mathbf{r})+g\left\vert \Psi\right\vert ^{2}\right)  \Psi, 
\label{3DGPE}
\end{equation}
where $g=4\pi\hbar^{2}a/m$ is the 
interaction strengh determined by the positive \textsl{s}-wave scattering 
length $a$ and the bosonic mass $m$, $V(\mathbf{\mathbf{r}})=m 
\omega_\perp^2 r_\perp^2/2$ is an external, cylindrically symmetric, harmonic
potential in the transverse coordinate $r_\perp^2=x^2+y^2$, and the 
condensate particle number $N$ follows 
from normalization $N=\int d\mathbf{r}^3 \left\vert \Psi\right\vert ^{2}$.

Our starting point is the search for stationary solutions to Eq.~(\ref{3DGPE}), 
$\Psi(\mathbf{r},t)=e^{-i\mu t/\hbar}\psi(\mathbf{r})$, with chemical potential 
$\mu$, having the form of planar kinks across the axial coordinate $z$. This 
task has been carried out numerically because no analytical solution is known 
for the 3D Gross-Pitaevskii equation (\ref{3DGPE}). Nevertheless, we have been 
guided by the asymptotic analytical solution proposed in Ref.~\cite{Mateo2014}
\begin{equation}
\psi(\mathbf{r})=\chi_{TF}({r}_\perp)\tanh(z/\xi(r_\perp)),
\label{TFDS}
\end{equation}
which is valid in the Thomas-Fermi regime, where 
$\chi_{TF}=\sqrt{\mu_\mathrm{loc}(r_\perp)/g}$ is the 
transverse ground state, $\mu_\mathrm{loc}(r_\perp) = \mu - V(r_\perp)$ is the local chemical potential, and a local healing length is defined by 
$\xi(\mathbf{r}_\perp)=\hbar/\sqrt{m\mu_\mathrm{loc}(r_\perp)}$.
Employing Eq.~(\ref{TFDS}) as initial ansatz, the numerical solution of Eq.\ 
(\ref{3DGPE}) is obtained without difficulty either using a Newton method or by 
imaginary time evolution.

The ansatz (\ref{TFDS}) also provides an excellent description of relevant 
properties of the kink, such as excitation energy or missing number of 
particles. We define the excitation energy of a soliton $\psi$, 
relative to the ground state $\psi_0$, by means of
\begin{equation}
E_s=E[\psi]-\mu N-(E[\psi_{0}]-\mu N_0) \, ,
\label{free_energy}
\end{equation}
with the energy defined by the functional
\begin{equation}
E[\psi]=\int d\mathbf{r}\left( \frac{\hbar^2}{2m}\left\vert\nabla 
\psi\right\vert^2 +
V(\mathbf{r})\left\vert \psi\right\vert^2 + \frac{g}{2}\left\vert 
\psi\right\vert^4\right) \, ,
\label{Energy}
\end{equation} 
Substituting the ansatz (\ref{TFDS}) into Eq.~(\ref{free_energy}) and neglecting 
the derivatives of $\chi_{TF}({r}_\perp)$ in Eq.~(\ref{Energy}), 
according to the Thomas-Fermi approximation,  we get the energy of a planar dark 
soliton confined by a transverse harmonic trapping as
\begin{equation}
\frac{E_s}{\hbar\omega_\perp}= 
\frac{4}{15}\frac{\tilde{\mu}^\frac{5}{2}}{\tilde{a}}\, ,
\label{EnergyDSansatz}
\end{equation} 
where the quantities with tilde are measured in the characteristic units of the 
trap, $\tilde{\mu}=\mu/\hbar\omega_\perp$, $\tilde{a}=a/a_\perp$, and 
$a_\perp=\sqrt{\hbar/m\omega_\perp}$ .
Normalization of Eq.~(\ref{TFDS}) gives the missing number of particles in the 
soliton $N_s=N-N_0 $ :
\begin{equation}
{N_s}= -\frac{2}{3}\frac{\tilde{\mu}^\frac{3}{2}}{\tilde{a}}\, ,
\label{particleDSansatz} 
\end{equation}
which can also be obtained from the relation 
\begin{equation}
\partial E_s / \partial\mu = -N_s \, .
\label{missing_number}
\end{equation}

\begin{figure}[htb]
\center
\includegraphics[width=0.7\linewidth]{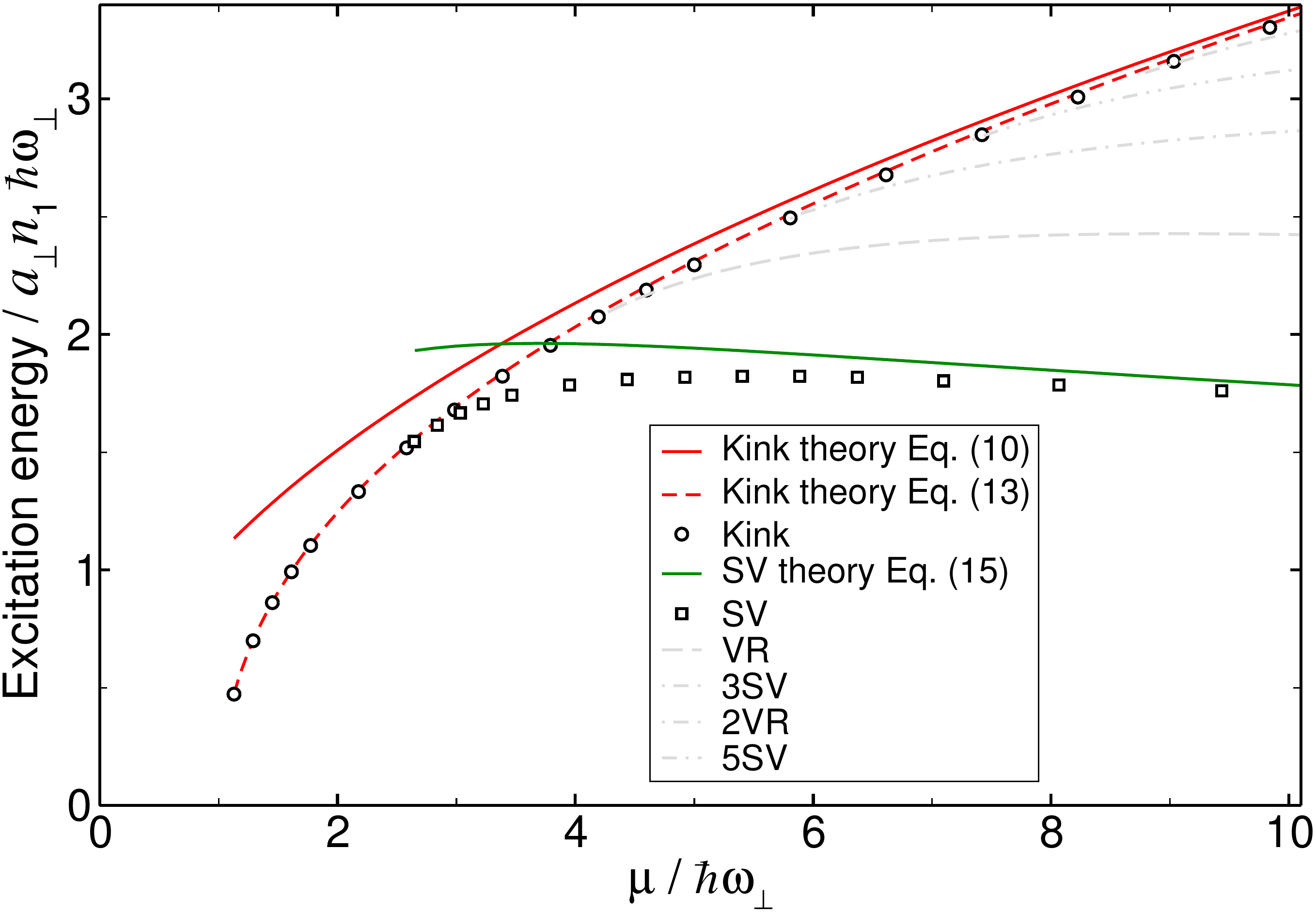}
  \vskip0.2cm
\includegraphics[width=0.7\linewidth]{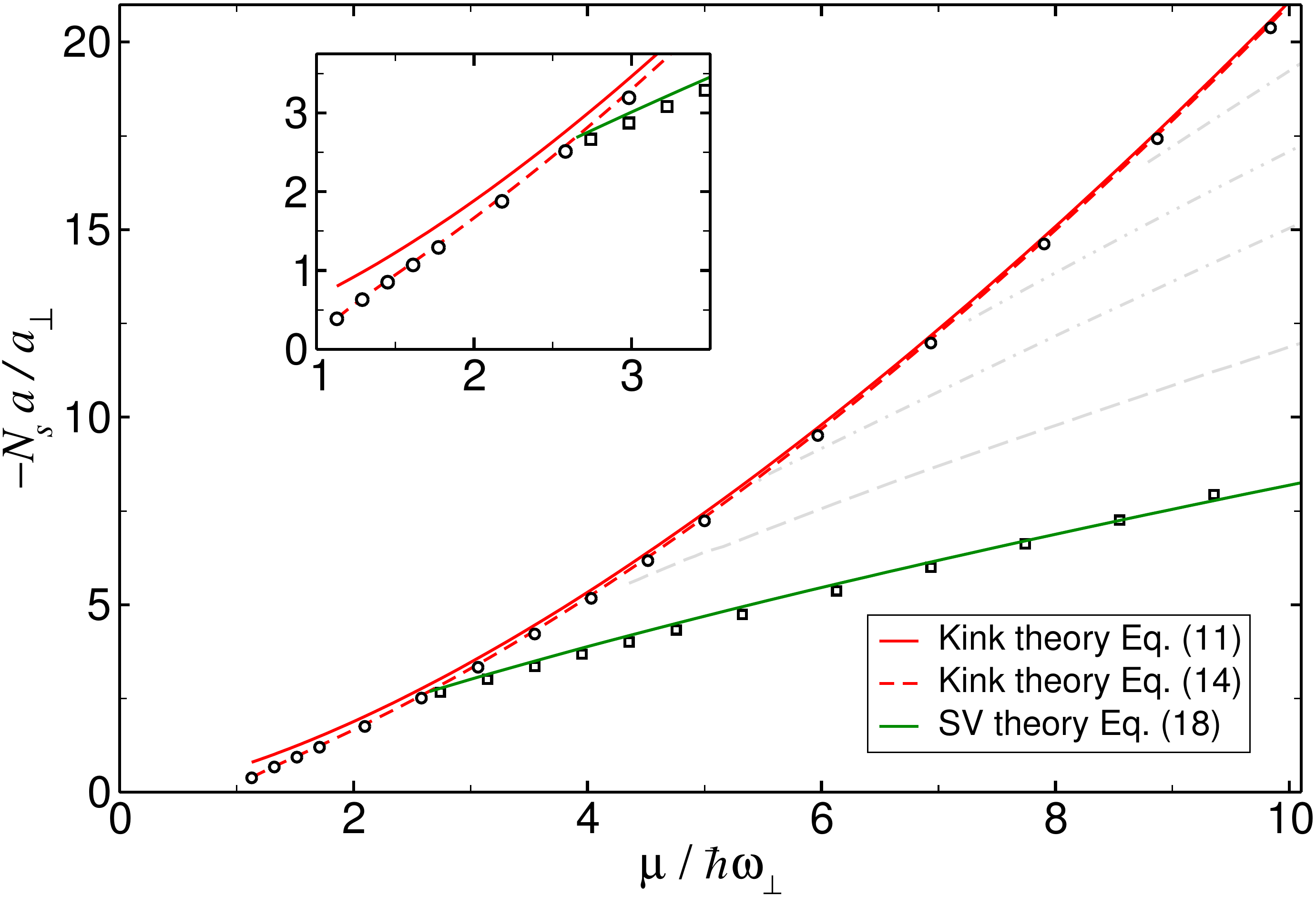}
\caption{Excitation energy (upper panel) and number of missing particles $N_s$ (lower panel) for a three dimensional solitary waves in an infinite cylindrical BEC with transverse trapping frequency $\omega_\perp$ as a function of chemical potential $\mu$. Results from fully numerical solutions of Eq.\ 
(\ref{3DGPE}) are shown as circles for the kink (dark soliton) and squares for the solitonic vortex (SV).
Formulas (\ref{EnergyDSansatz}) and (\ref{particleDSansatz}) based on the Thomas Fermi approximation for the kink are shown in a full red line. The improved approach of 
equations (\ref{EnergyDS}) and (\ref{particleDS}) is also shown (dashed red line) for comparison. The analytical approximations of equations \eqref{eq:ESV} and \eqref{eq:NSV} for the solitonic vortex are shown as a full green line.
In addition, grey lines show numerical results for
other stationary Chladni solitons: single vortex ring (VR), triple solitonic vortex (3SV),
double vortex ring (2VR), and quintuple solitonic vortex (5SV).}
\label{TFDS_comparison}
\end{figure}

Figure \ref{TFDS_comparison} shows a comparison of the analytical 
expressions (\ref{EnergyDSansatz}) and (\ref{particleDSansatz}) derived from 
the ansatz (\ref{TFDS}), and  the exact numerical solution to Eq.\ 
(\ref{3DGPE}). As expected, the agreement improves with increasing values of the 
chemical potential, and for $\tilde{\mu}>10$ errors are below $1\%$. The 
preceding analysis shows that for a harmonically trapped kink, given 
a chemical potential in the trap units, the parameters $\tilde{a} \tilde{E_s}$ 
and $\tilde{a} N_s$ are fixed, as reflected in equations (\ref{EnergyDSansatz}) 
and (\ref{particleDSansatz}).

A further improvement can easily be introduced in previous expressions for 
average quantities. Following Ref.~\cite{Mateo2007}, by properly 
incorporating the zero point energy of the harmonic oscillator in the 
calculations we obtain the improved expressions  
\begin{equation}
\frac{E_s}{\hbar\omega_\perp}= 
\frac{4}{15}\frac{(\tilde{\mu}-1)^\frac{5}{2}}{\tilde{a}}+
\frac{2}{3}\frac{(\tilde {\mu}-1)^\frac{3}{2}}{\tilde{a}}\, ,
\label{EnergyDS}
\end{equation} 
and
\begin{equation}
{N_s}= -\frac{2}{3}\frac{(\tilde{\mu}-1)^\frac{3}{2}}{\tilde{a}}-
\frac{(\tilde{\mu}-1)^\frac{1}{2}}{\tilde{a}}\, ,
\label{particleDS}
\end{equation} 
which have the correct limits both in the Thomas-Fermi and in the
quasi-onedimensional regimes. They also interpolate in between with very good 
accuracy as can be seen in Fig.~\ref{TFDS_comparison}.

\subsection{Solitonic vortex and vortex ring}\label{sec:SV}


We have determined the inertial and physical masses for solitary waves by computing the energy of the fully numerical solution of the 3D Gross-Pitaevskii equation \eqref{3DGPE} and taking numerical derivatives with respect to chemical potential and velocity near the stationary point. Figure  \ref{Fig2} reports the resulting mass ratios for dark solitons (red curve), solitonic vortices (green curve with open
squares), and vortex rings (blue curve with open circles) in 3D condensates
confined by isotropic radial harmonic traps. For the kink, one can observe
discontinuities arising at the bifurcation points of $(p,0)$ modes, 
corresponding to vortex rings. As will be explained in later sections, at these
points the kink solutions exist only for the static dark
soliton, and moving vortex rings emerge with the same value of the chemical
potential.

Approximate formulas for the solitonic vortex properties from hydrodynamic theory in logarithmic accuracy appeared in Ref.~\cite{Ku2014}. The energy expression for the stationary solitonic vortex in a BEC reads
\begin{eqnarray}\nonumber \label{eq:ESV}
E_{SV} &= \frac{\pi \hbar^2 m}{m_B^2} n_2 \ln\frac{R_\perp}{\xi} \\
& = \frac{\sqrt{2}}{3} \tilde{\mu}^{\frac{3}{2}}\ln(2\tilde{\mu}) \frac{a_\perp}{a}\hbar \omega_\perp\\
&=\frac{4\sqrt{2}}{3 \sqrt{\tilde{\mu}} }\ln(2\tilde{\mu}) a_\perp n_1 \hbar \omega_\perp ,
\end{eqnarray}
and the Thomas Fermi approximation has been used for the one- and two-dimensional
densities, in particular $4 a n_1 = \tilde{\mu}^2$.
The missing particle number is obtained by differentiation
\begin{eqnarray}
N_{SV} &
= - \frac{m}{m_B} \pi n_2 \xi^2\left(\frac{2\gamma+1}{\gamma} \ln\frac{R_\perp}{\xi} + \frac{2}{\gamma}\right)\\ \label{eq:NSV}
& = - \frac{\sqrt{2}}{2}\frac{a_\perp}{a} \sqrt{\tilde{\mu}}\left(\ln (2\tilde{\mu}) +\frac{2}{3}\right) ,
\end{eqnarray}
where $\gamma = \frac{\mu}{n}\frac{\partial n}{\partial \mu}$ is a polytropic index characterising the equation of state, which evaluates to $\gamma = 1$ for the case of a BEC.
For the inertial mass the following expression was obtained
\begin{eqnarray}
\Mi^{SV} &= -\frac{4 \pi}{2\gamma+1} \frac{n_2 R_\perp^2}{\ln\frac{R_\perp}{\xi}} m \\
&= -\frac{8\sqrt{2}}{9} \frac{\tilde{\mu}^{\frac{5}{2}}}{\ln(2\tilde{\mu})} \frac{a_\perp}{a} m .
\end{eqnarray}

\begin{figure}[tb]
\center
\includegraphics[width=0.7\linewidth]{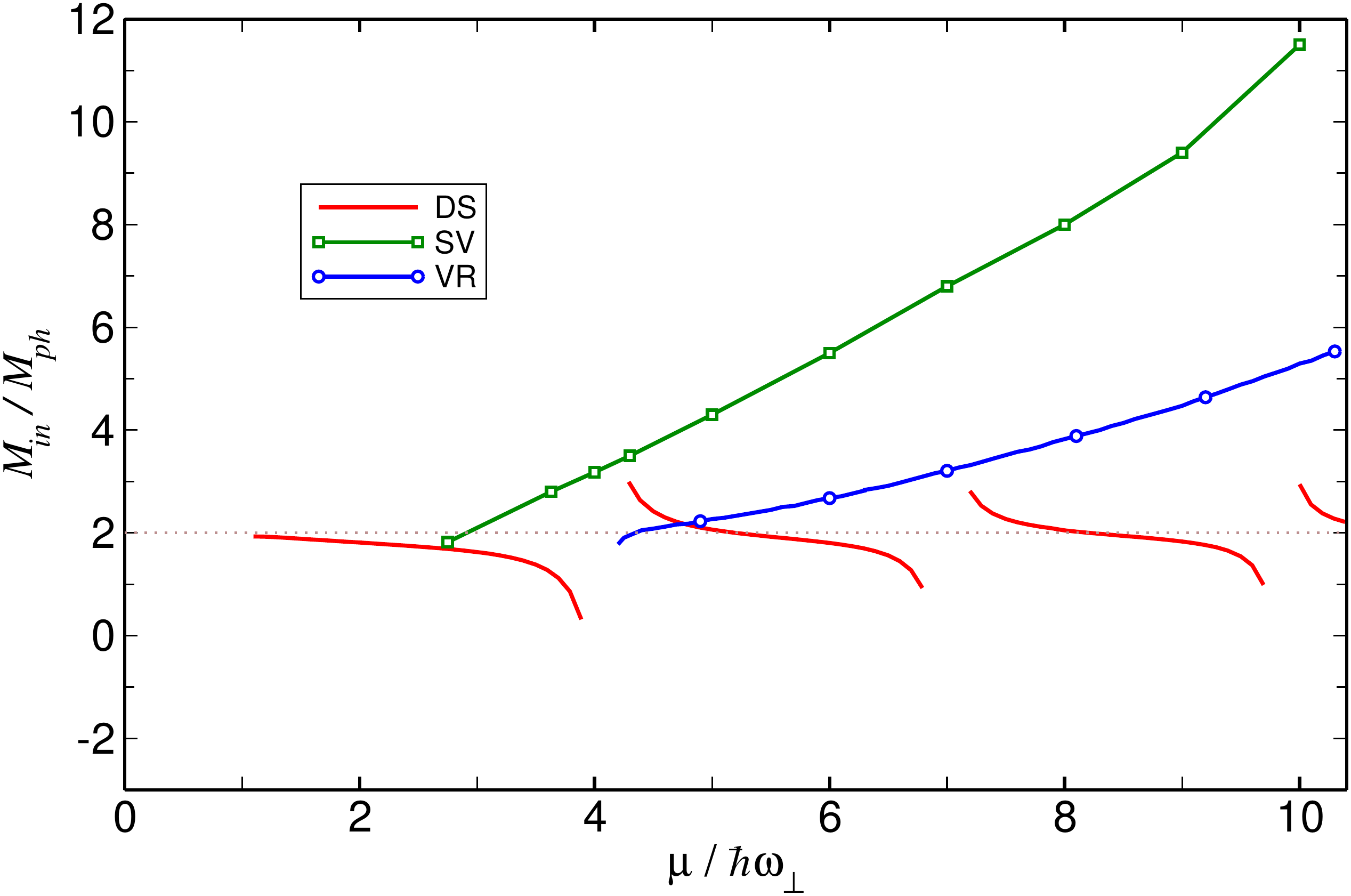}
\caption{ The ratio of inertial to physical mass as a function of the chemical potential $\mu$ for kinks (DS, red line), solitonic vortices (SV, green line with open square), and vortex rings (VR, blue line with open circles) in a cylindrical BEC with isotropic transverse harmonic trapping with frequency $\omega_\perp$.
}
\label{Fig2}
\end{figure}

\section{Snaking instability of the dark soliton}\label{sec:SI}

\subsection{Bogoliubov stability analysis}

After finding stationary kink solutions to Gross-Pitaevskii equation (\ref{3DGPE}), 
we look for elementary excitations $\{u(\mathbf{r}),v(\mathbf{r})\}$ with 
angular frequency $\omega$ around every equilibrium state $\psi$ with 
chemical potential $\mu$. The perturbed state can be written as 
$\Psi(\mathbf{r},t)=e^{-i\mu t/\hbar}\left[\psi+\sum_\omega(u \,e^{-i\omega t}+ 
v^* e^{i\omega t})\right]$, and the excitation modes are the solutions to the 
linear Bogoliubov equations, 
\begin{subequations}
\label{Bog} 
\begin{align}
\left( H_0-\mu + 2g\vert\psi\vert ^{2}\right) u +
g\psi^{2} v &= \hbar\omega u \; ,
\\ 
-g{\psi^*}^2 u -\left( H_0-\mu + 2g\vert\psi\vert ^{2}\right) v
 &= \hbar\omega v\; ,
\end{align}
\end{subequations}
where $H_0=-\hbar^{2}\nabla^{2}/2m+V(\mathbf{r}_\perp)$. Dynamical 
instabilities are related to  solutions to Eq.~(\ref{Bog}) 
with complex frequencies $\omega$, where the imaginary part of  $\omega$ is a rate of exponential growth of the corresponding unstable mode. The inverse of the imaginary part of the unstable frequency can thus be interpreted as the lifetime of the particular mode. Below we will report fully numerical solutions of Eq.~\eqref{Bog} for the dark soliton and Chladni solitons. Here we will proceed with finding analytical solutions to the Bogoliubov equations for the dark soliton based on the Thomas Fermi approximation.

\subsection{Approximate separation of variables}
For a real-valued stationary state, as it is the case of the dark soliton or kink of
 Eq. (\ref{TFDS}), the Bogoliubov equations (\ref{Bog}) can be transformed using $f_\pm(\mathbf{r})=u(\mathbf{r})\pm 
v(\mathbf{r})$ 
and $g_\pm=(2\pm 1)g$
 into
\begin{equation}
\left( H_0-\mu + g_\pm \psi^2\right) f_\pm = \hbar\omega f_\mp \; .
\label{BogReal}
\end{equation}
For $\omega=0$ the two equations decouple. This was the problem solved in Ref.~\cite{Mateo2014} in order to find the bifurcation points of Chladni solitons from the dark soliton. Here, we are interested in the more general problem of finding solutions to Eq.~\eqref{BogReal} with finite imaginary or complex values of $\omega$. Aiming at an approximate separation of transverse and longitudinal degrees of freedom we introduce the rescaled variable $\bar{z}=z/\xi(r_\perp)$.
For the dark soliton state of  Eq.~(\ref{TFDS})  we can write $g\psi^2 = \mu_\mathrm{loc}(r_\perp) \tanh^2 \bar{z}$, where  $\mu_\mathrm{loc}=\mu-V(r_\perp)$ and may thus rewrite the Bogoliubov equation \eqref{BogReal} as
\begin{align}
\left(-\frac{\hbar^2}{2m}\nabla_\perp^2 + \mu_\mathrm{loc} A_\pm \right)  f_\pm = \hbar\omega f_\mp ,
\end{align}
with
\begin{subequations}
\begin{align}
A_-&= -\frac{1}{2} \frac{\partial^2}{\partial \bar{z}^2}- 1+\tanh^2 \bar{z} ,\\
A_+&= -\frac{1}{2} \frac{\partial^2}{\partial \bar{z}^2}- 1+3\tanh^2 \bar{z} .
\end{align}
\end{subequations}
The operators $A_\pm$ are both Hamilton operators of one-dimensional Schr\"odinger equations with a shifted Rosen-Morse potential, whose eigenfunctions are known analytically \cite{Rosen1932}. Here we use the localised ground states of the respective operators as a restricted basis for the $z$-dependence of $f_\pm$, since we expect the unstable modes to be localised near the dark soliton plane. The ground state eigenfunction of $A_+$ is $\varphi_+^0 (\bar{z})= \frac{\sqrt{3}}{2}\mathrm{sech}^2 \bar{z}$ with eigenvalue $0$. It corresponds to the well-known Goldstone mode of translation of the dark soliton in $z$ direction. This does not constitute an  instability in itself but the mode will be relevant for constructing the decaying modes with imaginary omega. The operator $A_-$ has the ground state wave function $\varphi_-^{-1/2}(\bar{z}) = \frac{1}{\sqrt{2}}\mathrm{sech}\,\bar{z}$ with eigenvalue $-\frac{1}{2}$. It is this mode that is responsible for the existence of unstable Bogoliubov modes.

The $z$ dependence can now be removed from the Bogoliubov equation \eqref{BogReal} by starting from the ansatz $(f_+,f_-)^t =  \chi_+(x,t) (\varphi_+^0,0)^t + \chi_-(x,y) (0,\varphi_-^{-1/2})^t $. Ignoring any $x,y$ derivatives of the functions $\varphi_\pm(\bar{z})$ and projecting onto the respective ground  states by multiplying from the left with $(\varphi_+^0,0)$ and $(0,\varphi_-^{-1/2})$, 
and integrating over $\bar{z}$, we obtain the matrix equation 
\begin{align} \label{eq:BogTrans}
\left( \begin{array}{cc}
-\frac{\hbar^2}{2m}\nabla_\perp^2& - \hbar\omega \zeta\\
 -\hbar\omega \zeta& -\frac{\hbar^2}{2m}\nabla_\perp^2  - \frac{1}{2} \mu_\mathrm{loc} 
 \end{array} \right)
\left(\begin{array}{c}
\chi_+\\
\chi_-
\end{array}\right)
 =0 ,
\end{align}
where $\zeta=\int \varphi_-^{-1/2} \varphi_+^0\,d\bar{z} = \frac{\pi}{4}\sqrt{\frac{2}{3}}\approx 0.962$ is a numerical constant close to one.

\subsection{Homogeneous background}
The transverse Bogoliubov equation \eqref{eq:BogTrans} is easily solved in the absence of a transverse trapping potential, where $\mu_\mathrm{loc} = \mu = \frac{\hbar^2}{m\xi^2}=\mathrm{const}$, with plane wave solutions, e.g., $\chi_\pm \propto \exp(i q x)$. For the unstable eigenvalues we find
\begin{align} \label{eq:homBog}
\omega^\mathrm{hom} = i q \frac{\sqrt{\mu}}{2c \sqrt{m}}  \sqrt{1-q^2 \xi^2} .
\end{align}
For small $q$ the growth rate $\mathrm{Im}(\omega)$ grows linearly with the slope $\frac{\sqrt{\mu}}{2\zeta \sqrt{m}} \approx 0.520  \frac{\sqrt{\mu}}{\sqrt{m}}$ as is demanded by general hydrodynamic arguments \cite{Kamchatnov2008}.
Although being approximate due to the restricted basis expansion of the $z$ dependence, this result compares very well with the previously obtained ones in Refs.~\cite{Kuznetsov1988,Muryshev1999,Kamchatnov2008} , where the exact slope for the Gross-Pitaevskii equation is $\frac{\sqrt{\mu}}{\sqrt{3 m}} \approx0.577 \frac{\sqrt{\mu}}{\sqrt{m}}$. The snaking instability is suppressed and eigenvalues become real-valued for wave numbers larger than $q_\mathrm{crit}=1/\xi$, which is the exact value. For intermediate values  $0<q<1/\xi$ the growth rate has previously only been obtained numerically, and Eq.~\eqref{eq:homBog} reproduces the results of Refs.~\cite{Kuznetsov1988,Muryshev1999} very closely.

\subsection{Harmonically trapped kink state}

We now proceed with solving the transverse Bogoliubov equation \eqref{eq:BogTrans} for  an isotropic transverse trapping potential with $\mu_\mathrm{loc}(r_\perp) = \mu - \frac{1}{2}m\omega_\perp^2 r_\perp^2$.
Assuming an azimuthal dependence $\propto \cos(l\theta)$  with the quantum number $l=0,1,2,\ldots$ reduces Eq.~\eqref{eq:BogTrans} to a set of ordinary differential equations in the radial coordinate $r_\perp$. It is now convenient to move to harmonic oscillator units. Introducing the rescaled radial coordinate $\tilde{r}_\perp = r_\perp/a_\perp$ and dividing the equation by $\hbar \omega_\perp$, we obtain
\begin{align} \label{eq:BogHo}
\left( \begin{array}{cc}
 H_1^l  & - \tilde{\omega} \zeta\\
- \tilde{\omega} \zeta & H_2^l- \frac{\tilde{\mu}}{2}
 \end{array} \right)
\left(\begin{array}{c}
\chi_+^l\\
\chi_-^l
\end{array}\right)
 =0 ,
\end{align}
where 
\begin{align}
H_1^l &= -\frac{1}{2}\left(\frac{\partial^2}{\partial \tilde{r}_\perp^2} + \frac{1}{\tilde{r}_\perp}\frac{\partial}{\partial \tilde{r}_\perp} - \frac{l^2}{\tilde{r}_\perp^2} \right), \\
H_2^l &= H_1^l + \frac{1}{4}\tilde{r}_\perp^2 ,
\end{align}
where $H_1^l$ represents a two-dimensional Laplacian and $H_2^l$ is the Hamiltonian of a two-dimensional harmonic oscillator with weakened trap potential compared to the one experienced by the atoms. Equation \eqref{eq:BogHo} can also  be rewritten in the form
\begin{align} \label{eq:BogHoSquared}
H_1^l \left(H_2^l- \frac{\tilde{\mu}}{2}\right) \chi_-^l = \tilde{\omega}^2 \zeta^2 \chi_-^l .
\end{align}
Even though this represents a non-hermitian eigenvalue problem, we have only found real eigenvalues $\tilde{\omega}^2 \zeta^2$ in numerical investigations.

\subsection{Chladni soliton bifurcation points}
Solutions of Eq.~\eqref{eq:BogHo} with $\tilde{\omega}=0$ have special significance as they indicate the transition of a specific mode from representing an instability $\tilde{\omega}^2<0$ to a stable small amplitude oscillation $\tilde{\omega}^2>0$. At the same time they indicate a bifurcation of the stationary nonlinear solutions of the Gross-Pitaevskii equation (\ref{3DGPE}) and here relate to the branch-off points for Chladni solitons. The solutions are found in terms of the scaled Fock-Darwin radial eigenfunctions $\chi_p^l(\tilde{r}_\perp) = 2^{-\frac{1}{4}}R_p^l(2^{-\frac{1}{4}}\tilde{r}_\perp)$ of the 2D harmonic oscillator Hamiltonian $H_2^l$ with eigenvalues $\epsilon_p^l = (2p + l +1)/\sqrt{2}$ and
\begin{align}
R_p^l(r) = \sqrt{\frac{2 p!}{(p+l)!}}r^l L_p^l\left({r^2}\right) e^{-\frac{r^2}{2}} ,
\end{align}
where $L_p^l(x)$ is the generalised Laguerre polynomial. It is easily seen that $\chi_p^l$ solves the Bogliubov equation \eqref{eq:BogHoSquared} with $\tilde{\omega}=0$ when $\epsilon_p^l =\tilde{\mu}/2$, which translates into the condition
\begin{align} \label{eq:bif}
 \frac{\mu}{\hbar \omega_{\perp}} = \sqrt{2}({2p +l +1}) ,
\end{align}
for the bifurcation points of Chladni soliton solutions from the dark soliton, as found previously in Ref.~\cite{Mateo2014}.

\subsection{Finite growth rates}
For finite instability rates $\mathrm{Im}(\tilde{\omega})$ the eigenvalue equation  \eqref{eq:BogHoSquared} can be expanded in a basis of the normalised eigenfunctions $|p,l)\equiv \chi_p^l$ of $H_2^l$, which transforms it into a tridiagonal matrix eigenvalue equation 
\begin{align} \label{eq:MatrixEV}
\sum_{p'} B^l_{p,p'} v_{p'} = \tilde{\omega}^2 \zeta^2 v_p
\end{align}
with $B^l_{p,p'} =(p,l|H_1^l (H_2^l- \frac{\tilde{\mu}}{2})|p',l)=\int_0^\infty \tilde{r}_\perp\,d\tilde{r}_\perp \chi_p^l H_1^l (H_2^l- \frac{\tilde{\mu}}{2}) \chi_{p'}^l$. For the matrix elements we find
\begin{align}
B^l_{p,p}& = \frac{(2p+l+1)^2}{4} - \tilde{\mu} \frac{2p+l+1}{4\sqrt{2}},\\
B^l_{p,p-1}&= \frac{1}{4}[2(p-1)+l+1]\sqrt{p^2+lp}- \tilde{\mu}  \frac{\sqrt{p^2+lp}}{4\sqrt{2}},\\
B^l_{p-1,p}&= \frac{1}{4}(2p+l+1)\sqrt{p^2+lp}- \tilde{\mu}  \frac{\sqrt{p^2+lp}}{4\sqrt{2}},
\end{align}
and $B^l_{p,p'} =0$ for $|p-p'|>1$. The matrix is block-diagonal in the azimuthal quantum number $l$ due to the azimuthal symmetry of the problem. We have solved the corresponding eigenvalue equations numerically and have found the approximate asymptotic behaviour
\begin{align} \label{eq:asymptotics}
 \tilde{\omega}_n^l \sim  \frac{\tilde{\mu} - \sqrt{2} (2n+1)}{4 \zeta},
\end{align}
for large $\tilde{\mu}$ and values of the azimuthal quantum number $l>1$. All these eigenvalues have a zero crossing for a finite value of $\tilde{\mu}$ corresponding to Eq.~\eqref{eq:bif}, where $n=p$ can be identified. These results were obtained by diagonalising a truncated matrix $B^l_{p,p'}$ with $p,p'< p_c$.  Changing the cutoff value $p_c$ affects the large-$\tilde{\mu}$ regime but leaves the zero crossings for $n\ll p_c$ unaffected. The asymptotic behaviour reported above represents the limit of $p_c\to\infty$.

In the $l=0$ sector a special case occurs, where the $|0,0)$ state needs to be excluded from the basis in order to eliminate an unphysical unstable eigenvalue with $p=0$ that otherwise occurs in solving Eq.~\eqref{eq:MatrixEV}. The mode with $p=0, l=0$ corresponds to zero point motion of the kink, which is not captured correctly by the underlying Thomas Fermi approximation. Indeed, no unstable mode with these quantum numbers is found in the full numerical solution of the Bogoliubov equations. Diagonalising Eq.~\eqref{eq:MatrixEV}  in a truncated basis with $0< p<p_c$ produces the asymptotic behaviour 
\begin{align} \label{eq:asymptoticslzero}
 \tilde{\omega}_n^{l=0} \sim  \frac{\tilde{\mu} - \sqrt{2} (2n-1)}{4 \zeta},
\end{align}
where still $n=p$ can be identified at the zero crossings of $\tilde{\omega}$.

The results of the truncated eigenvalue problem \eqref{eq:MatrixEV} with the asymptotic results \eqref{eq:asymptotics} and \eqref{eq:asymptoticslzero} closely resemble the full numerical results. 

%

\subsection{Fully numerical results}

\begin{figure}[tb]
\center
\includegraphics[width=0.75\linewidth]{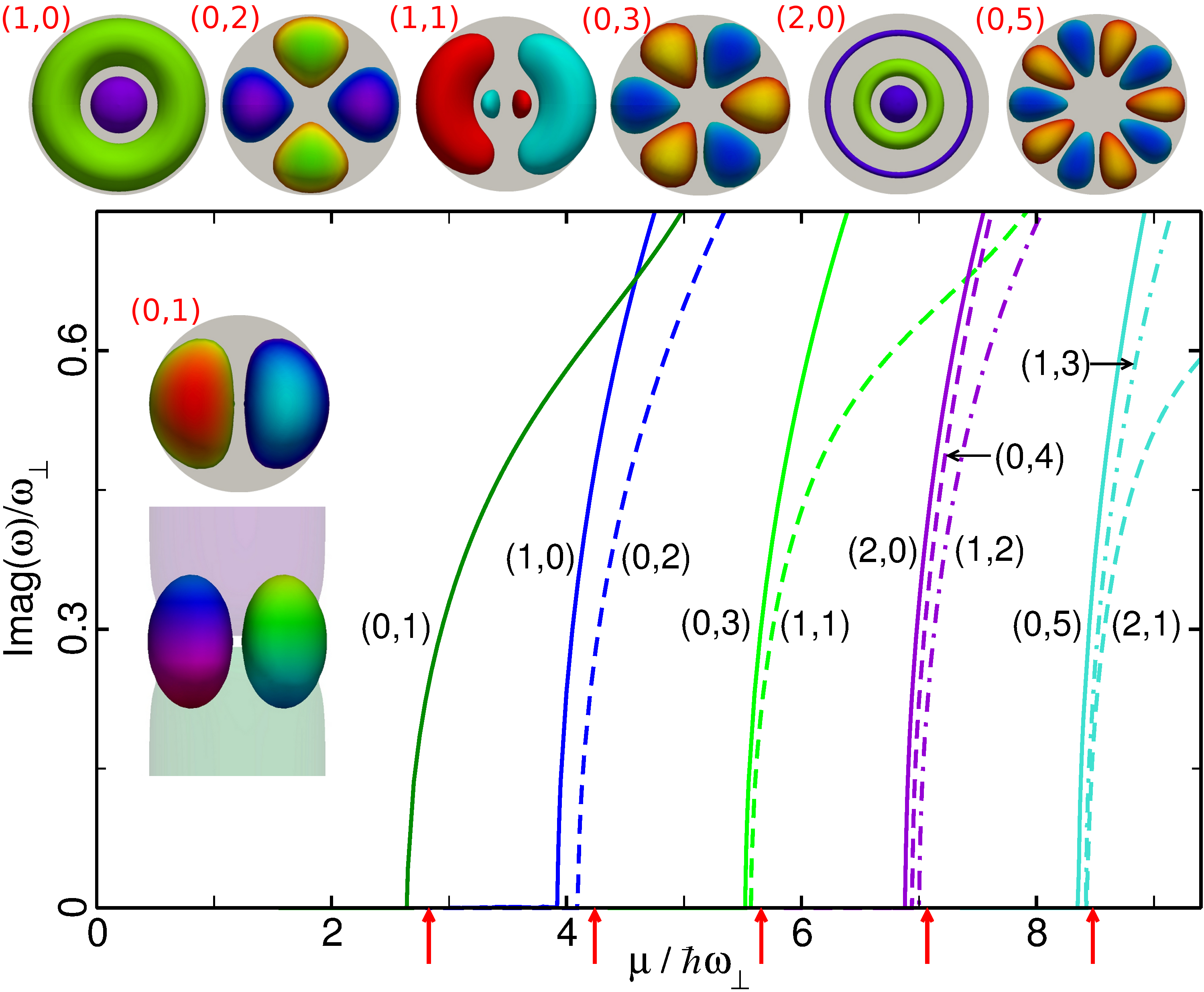}\\
\includegraphics[width=0.75\linewidth]{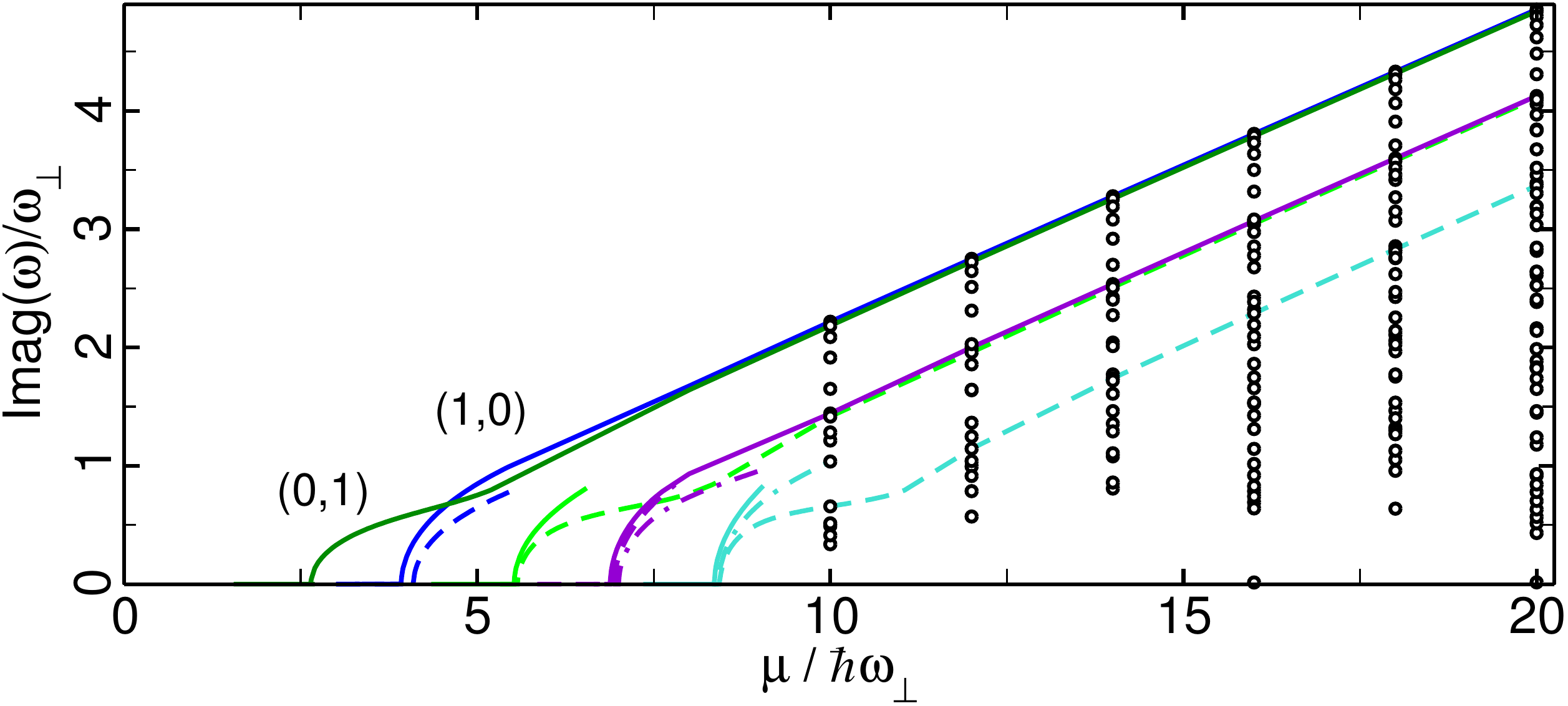}
\caption{Upper panel: Unstable frequencies $\omega$ and density isocontours at 
5$\%$ of maximum density (insets) of Bogoliubov modes, classified by 
their radial and angular quantum numbers $(p,l)$, responsible for the decay of the kink. The shaded background of the isocontours indicates the BEC density distribution.  
The axial view for the $(0,1)$ mode (bottom left inset) clearly demonstrates the axial localisation of the mode function. 
All the mode functions have been generated close to their respective
bifurcation points. The arrows below the horizontal axis indicate the bifurcation points according to the 
analytical prediction (\ref{eq:bif}).
Lower panel: Following the unstable mode frequencies to larger interaction parameters $\mu/(\hbar \omega_\perp)$ demonstrates the linear growth.
}
\label{Fig2}
\end{figure}

We have also solved the Bogoliubov equation \eqref{Bog} numerically in full three dimensions without making use of the approximations discussed in the previous paragraphs. The results for the  
unstable modes and associated frequencies are 
collected as a function of the chemical potential of the kink in Fig.\ 
\ref{Fig2}. The insets represent axial views of phase-colored density 
isocontours of the excitation modes $(p,l)$ just after the appearance of 
bifurcation points. These modes present a structure of nodal lines, derived 
from the radial $p$ and azimuthal $l$ quantum numbers, characteristic of the 
linear excitations of the transverse trap. As can be seen in the lower left 
inset, the only one displaying a longitudinal view, they are strongly localized 
around the plane of the kink. Their emergence follow in a very good 
approximation the analytical prediction for bifurcations given by Eq. 
(\ref{eq:bif}), which are indicated by red arrows below the horizontal axis. As 
the interaction energy increases from the quasi-onedimensional configuration, 
where kink states are stable structures and generate only real frequency 
excitations, the first bifurcation point denoting the appearance of a complex 
frequency for the mode $|\,p=0,l=1\rangle$ comes into existence at 
$\mu_{0,1}=2.65\hbar\omega_\perp$, very close to the $2.8\hbar\omega_\perp$ 
value predicted by (\ref{eq:bif}). From this point on, kinks are unstable 
states, and further increase of the chemical potential is accompanied by the 
emergence of new bifurcation points grouped around the integer energy values 
$\epsilon_{pl}=2p+l+1$, with the characteristic degeneracy of the 
two-dimensional harmonic oscillator.

\section{Stationary Chladni solitons} \label{sec:Chladni}

\begin{figure}[htb]
\center
\begin{tabular}{@{}ccc@{}}
\includegraphics[width=3cm]{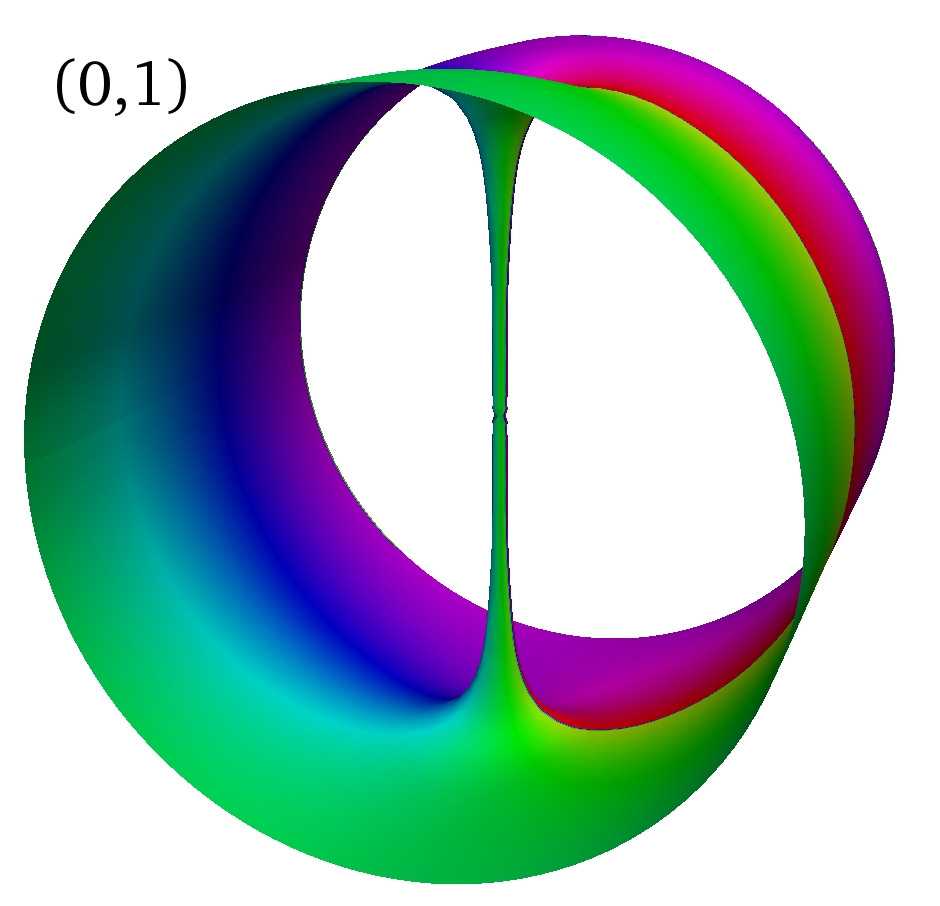} & 
\includegraphics[width=3cm]{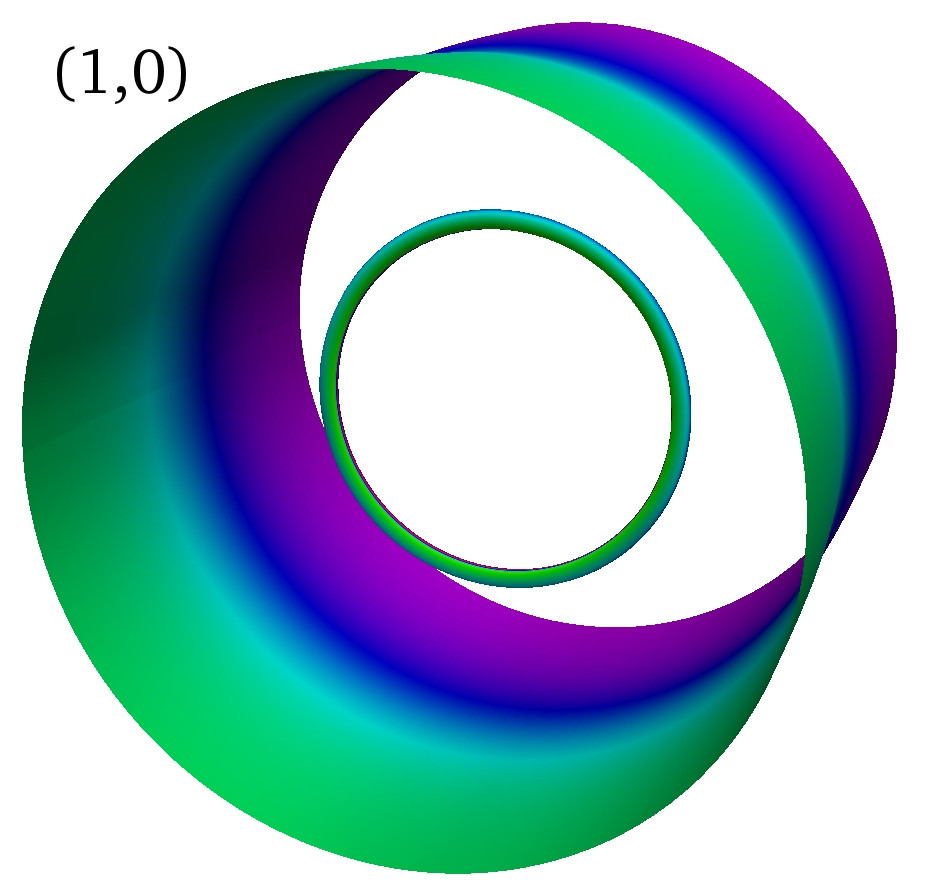} &
\includegraphics[width=3cm]{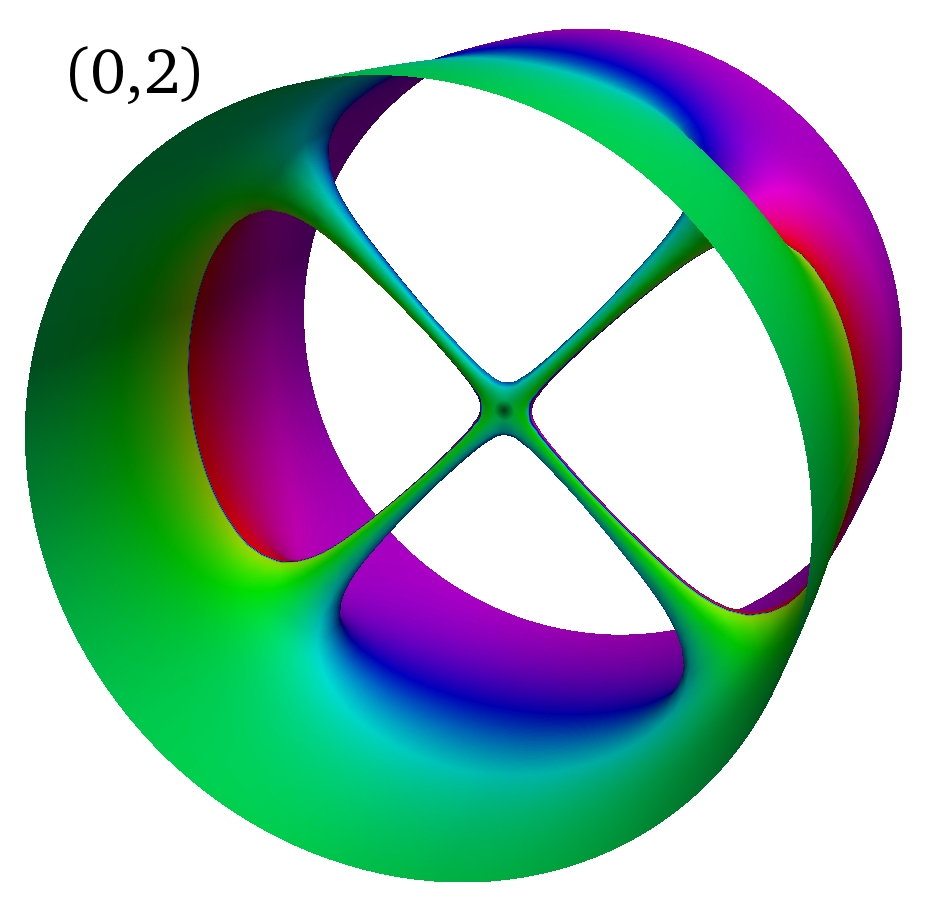}  \vspace{-0.15cm}\\ 
\includegraphics[width=3cm]{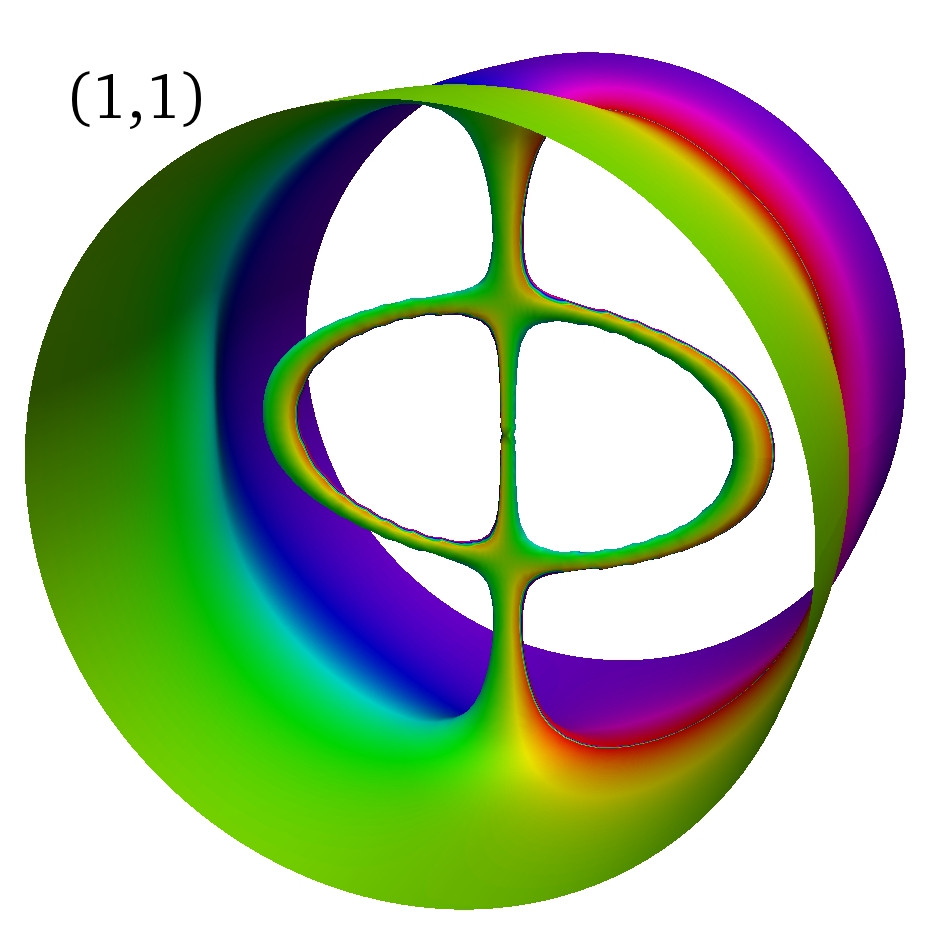} &
\includegraphics[width=3cm]{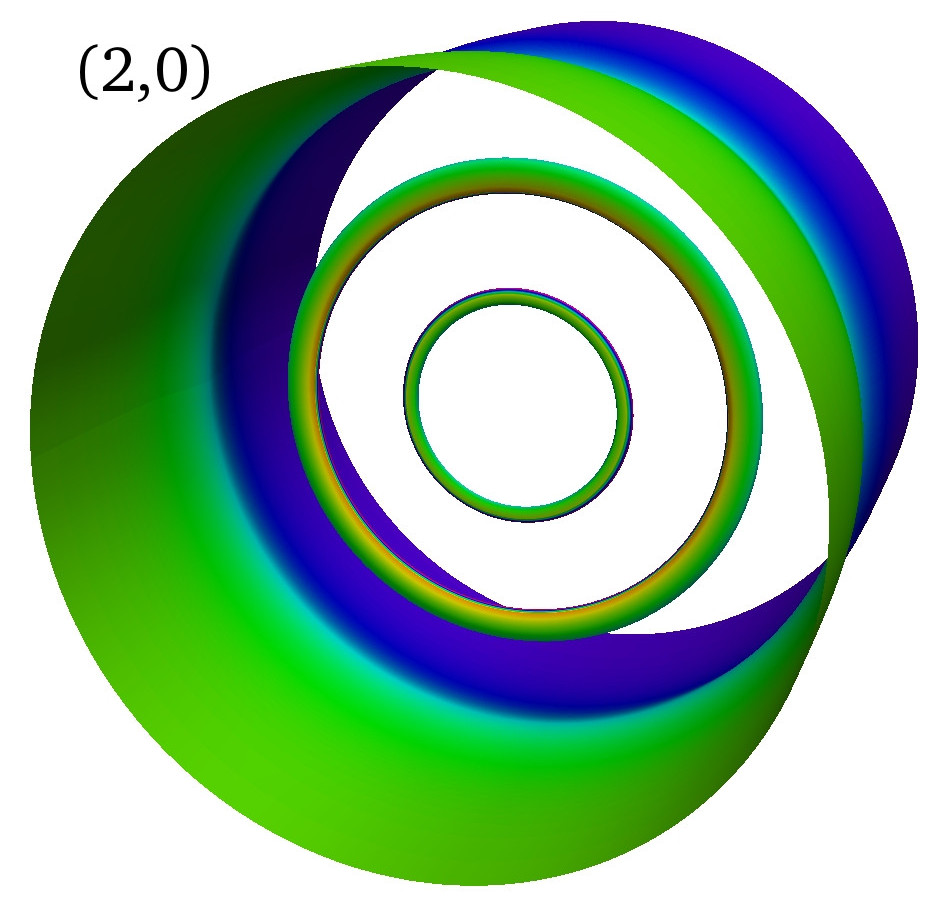} &
\includegraphics[width=3cm]{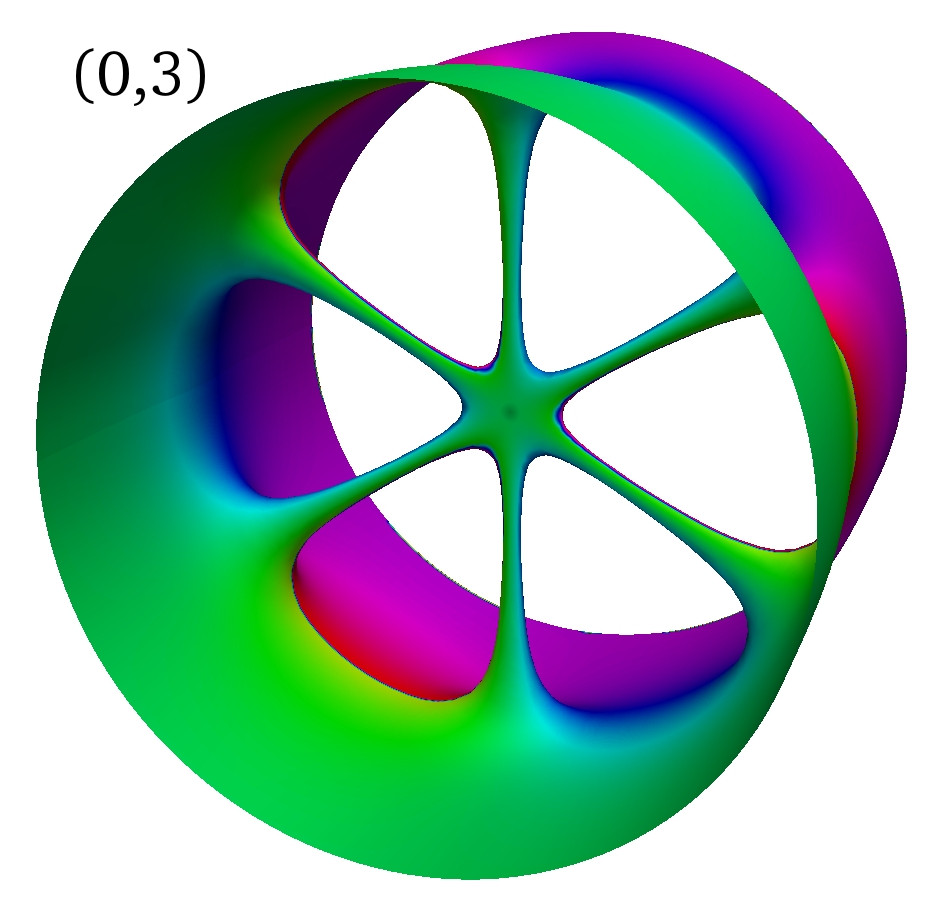} \vspace{-0.12cm}\\ 
\includegraphics[width=3cm]{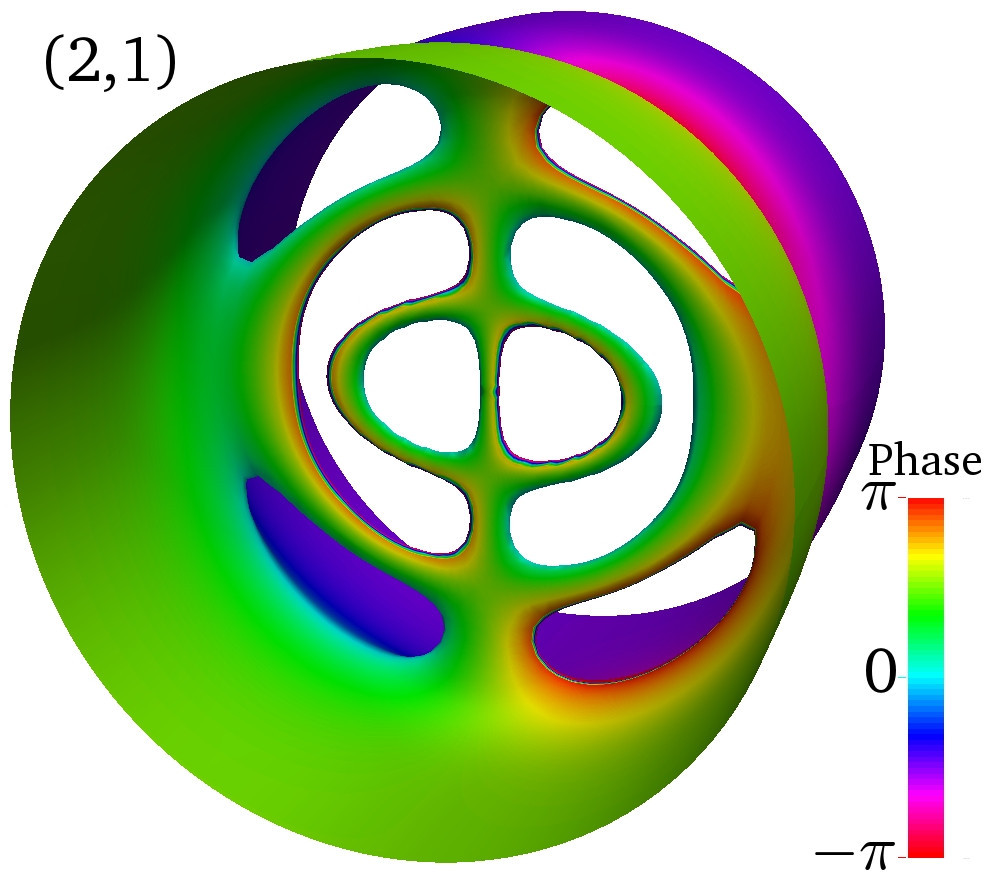} & 
\includegraphics[width=2.9cm]{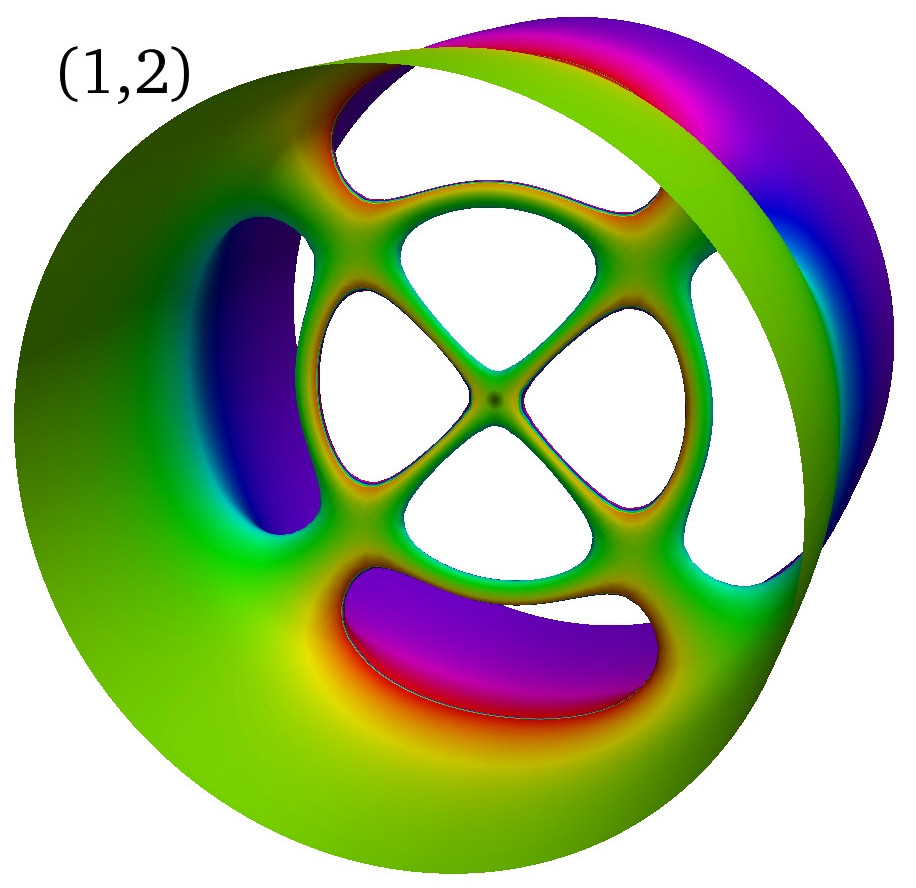} &
\includegraphics[width=3cm]{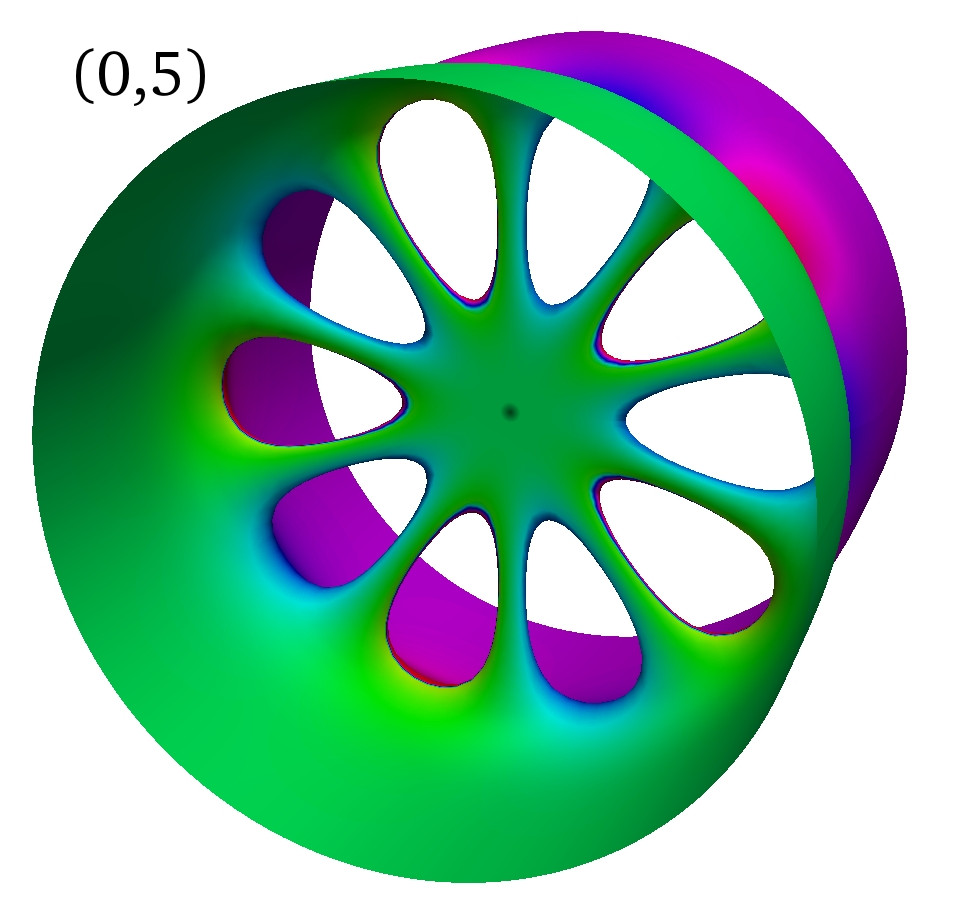} 
\end{tabular}
\caption{ Density isocontours (at 5 $\%$ of maximum density)
of static Chladni solitons with $\mu = 10  \,\hbar 
\omega_\perp$. The
cross-sections are $9\,a_\perp$ in width.}
\label{Fig3}
\end{figure}

Every unstable mode of the kink is associated with a stationary Chladni soliton. The zero crossings of the unstable mode frequencies in Fig.~\ref{Fig2} indicate bifurcation points of the Chladni solitons from the dark soliton. The sign patterns of the unstable mode functions at the bifurcation points $\chi_p^l(r_\perp) \cos(\theta)$ determine the direction of flow along or counter the $z$ direction and the nodal lines translate into vortex lines. Numerically obtained mode functions are also shown in Fig.~\ref{Fig2}. Increasing the nonlinearity parameter $\tilde{\mu} = \mu/(\hbar \omega_\perp)$ above the bifurcation point, we obtain a family of stationary Chladni solitons with the same structure and symmetries, as endowed by the unstable mode at the bifurcation point. A number of Chladni soliton solutions is shown in Fig.~\ref{Fig3}.
For instance, the first excited mode $|\,p=0,l=1\rangle$ generates the family 
of solitonic vortex states, while the next two unstable modes, $|\,1,0\rangle$ and 
$|\,0,2\rangle$, that branch off near $\mu_{1,0}=4\hbar\omega_\perp$ in Fig. \ref{Fig2}, generate the families of vortex rings and 
two crossed vortices states, respectively. 

The linear solutions of a general, isotropic or anisotropic, twodimensional harmonic oscillator can be 
written in terms of Hermite modes with cartesian symmetries instead of the Laguerre modes of cylindrical symmetry. The Hermite modes are characterised by the pair $(n_x,n_y)$ of
quantum numbers indicating the nodal lines along $x$ and $y$ directions. 
Corresponding Chladni solitons would be made of vortex lines in such rectangular patterns. Although we have found the stationary Chladni solitons bifurcating from the kink to all conform to the Laguerre type, one may still expect to find Hermite-type structures to be relevant in the decay process of the kink. In the linear regime, the Hermite 
modes $|n_x,n_y\rangle$ can be constructed as superposition of Laguerre modes 
$|p,l\rangle$ with degenerate energies $\epsilon_{n_x n_y}=\epsilon_{pl}$.
For example,$|n_x=0,n_y=2\rangle\propto |p=0,l=2\rangle- |p=1,l=0\rangle$, and 
$|n_x=1,n_y=2\rangle\propto |p=0,l=3\rangle-|p=1,l=1\rangle$. Because of 
the energy splitting of the bundle of linear degenerate states presented in 
Fig. \ref{Fig2}, the linear superposition mechanism is not acting in 
the nonlinear case. 
However, we have found that Hermite-like modes do emerge as 
nonlinear bifurcations at a second stage. In the first 
stage, families of stationary solitary waves bifurcate from the kink in close relation to its unstable Laguerre-like modes. Except for the 
solitonic vortex, all these families are made of unstable states, the 
unstable excitation modes of which can generate new solitary waves at a second stage. It 
is then when the new solitonic families turn out to be composed of Hermite-like  
modes $(n_x,n_y)$. The instabilities of Chladni solitons will be discussed in more detail in section \ref{sec:Stability}.

\section{Moving Chladni solitons} \label{sec:Moving}

\begin{figure}[htb]
\center
\includegraphics[width=0.6\linewidth]{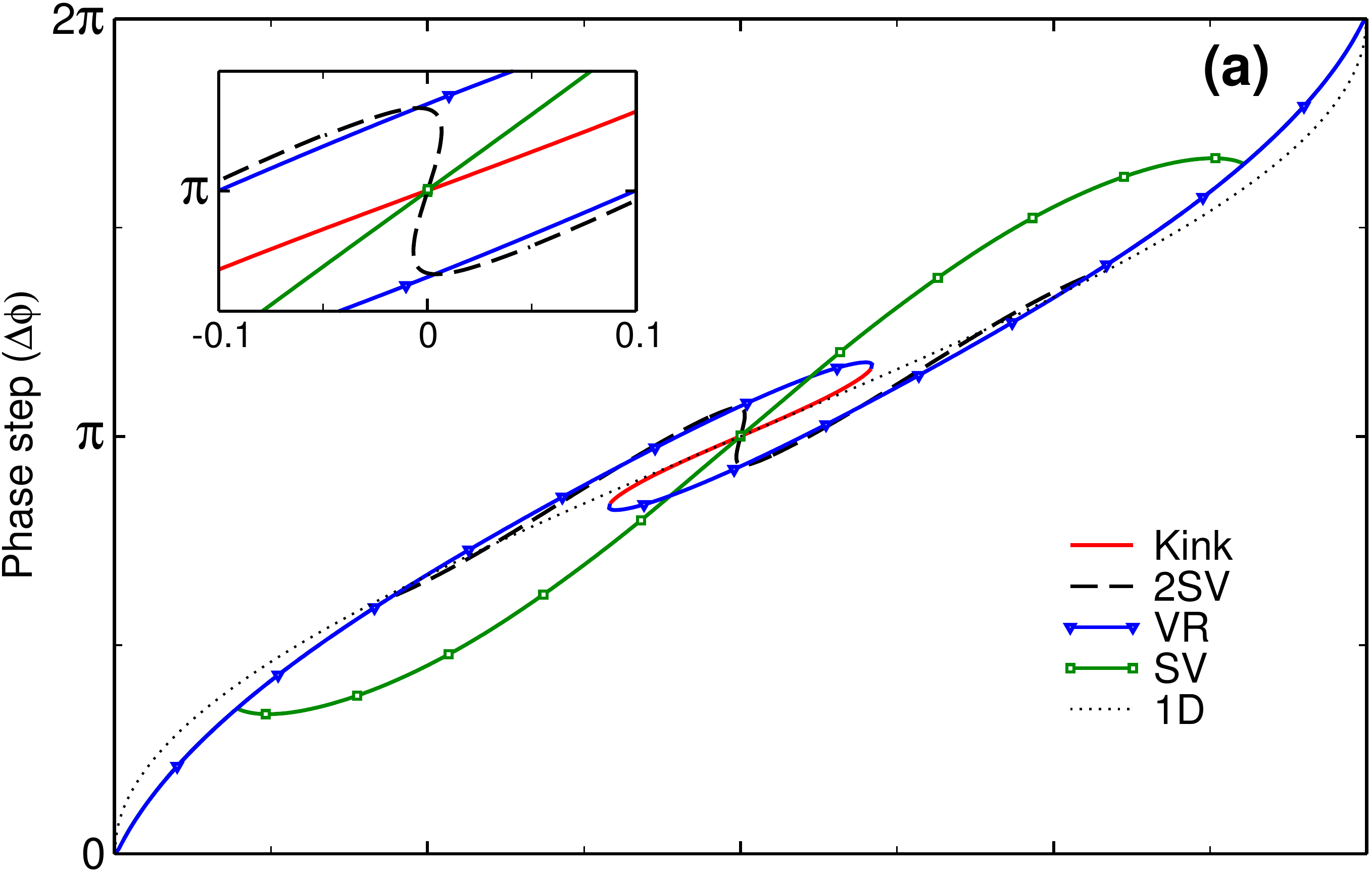}\\
\includegraphics[width=0.6\linewidth]{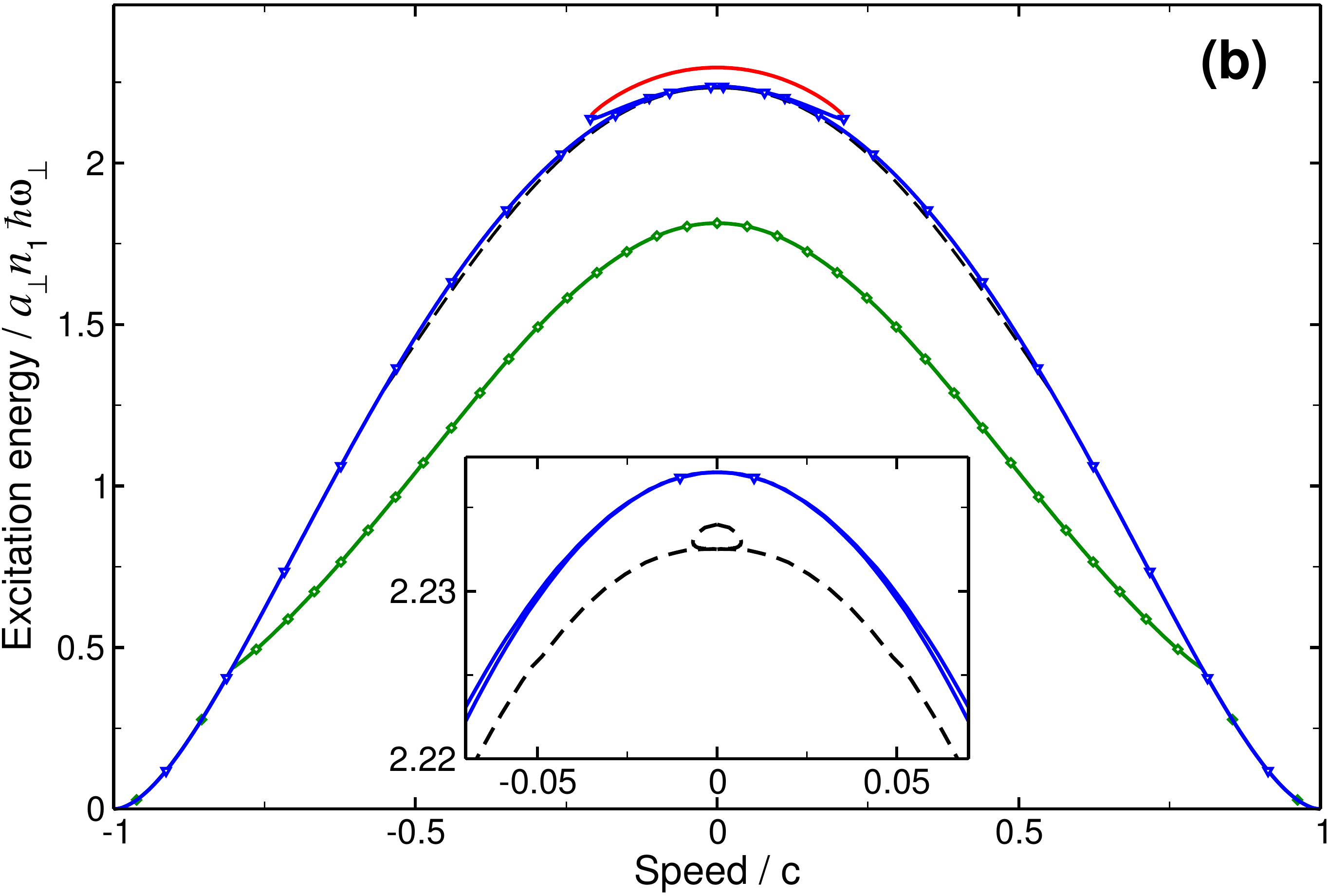}\\
\includegraphics[width=0.6\linewidth]{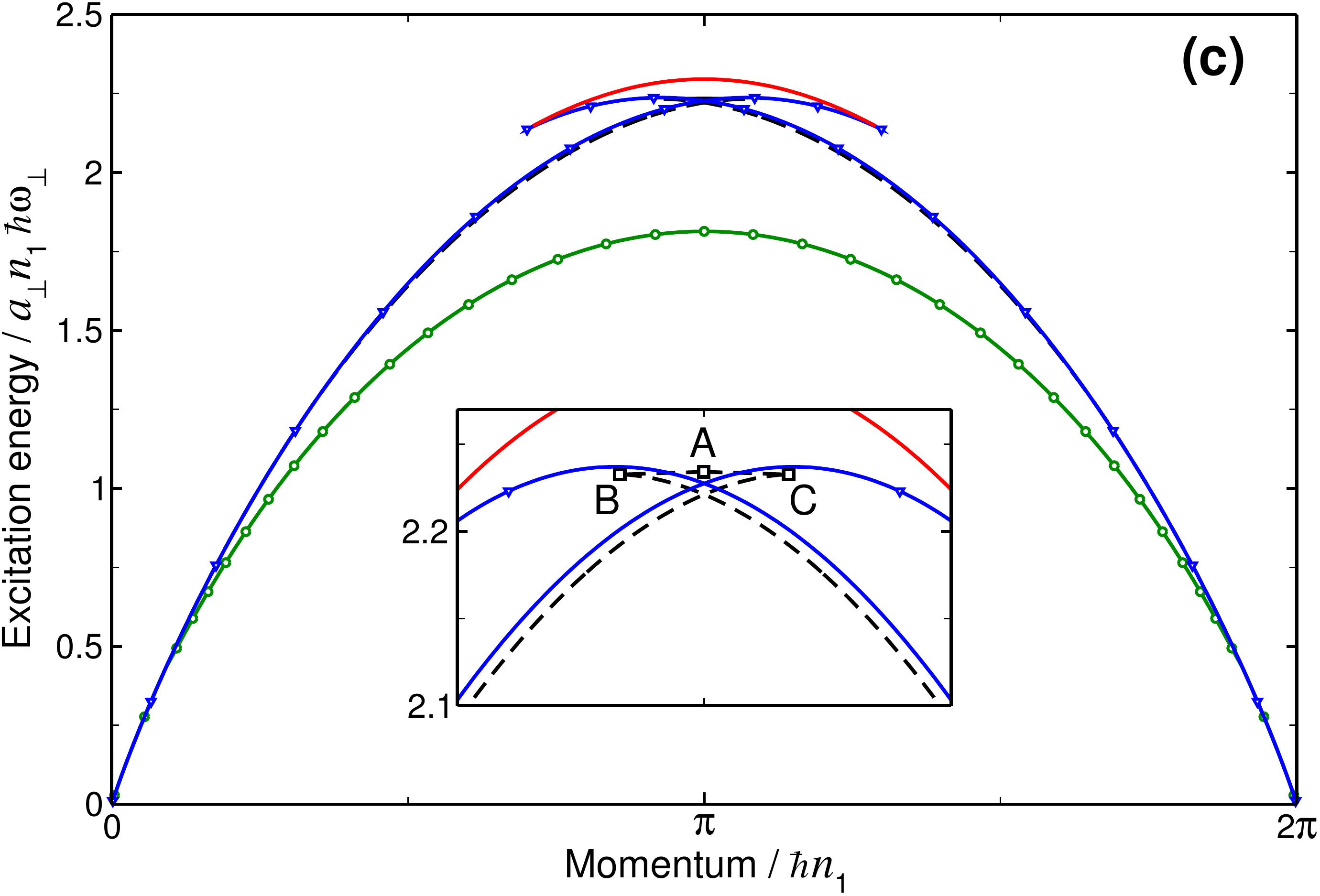}
\caption{ Characteristic quantities for solitary waves with chemical potential 
$\mu=5\hbar\omega_\perp$: axial phase step $\Delta\phi$ (a), and excitation 
energy $E_s$ (b) as a function of the axial velocity $v_z$ in units of the 
speed of sound $c$, and energy ($E_s$) vs.\ momentum dispersion relations (c).}
\label{Fig4}
\end{figure}
Up to now we have been looking at static solitons. In order to understand 
how the different families of solitonic states appear and connect, it is 
convenient to consider a more general picture of moving solitary waves.
Here, we extend upon the work of Komineas and Papanicolao, who already computed energy dispersion relations and phase steps for axisymmetric solitary waves (kinks and nested vortex rings) \cite{Komineas2002,Komineas2003a} and for the solitonic vortex \cite{Komineas2003}.

In order to construct the full dispersion relations for moving solitary waves, we numerically search for states $\Psi(\mathbf{r},t)=e^{-i(\mu' 
t+mv_z z)/\hbar}\psi(\mathbf{r})$, moving along the $z$-axis with a 
constant velocity $\mathbf{v_z}=(0,0,v_z)$, which are solutions of the 
stationary Gross-Pitaevskii equation for a co-moving reference frame:
\begin{eqnarray}
\frac{1}{2m}(-i\hbar\mathbf{\nabla}-m\mathbf{v_z})^2\psi+ 
V(\mathbf{r})\psi + g\vert \psi \vert^2 \psi=\mu'\psi \, ,
\label{vGPE}
\end{eqnarray}
where $\mu'=\mu+mv_z^2/2$ is the shifted chemical potential. Moving solitons 
have nonzero density dips associated to reduced (smaller than $\pi$) phase 
jumps, and their velocities are limited by the speed of sound $c$, at which 
solitonic states become linear sound excitations. In order to calculate the speed of sound, which will be used as velocity unit in what follows, we will make use 
of the analytical expression given in Ref.~\cite{Mateo2006}:
\begin{equation}
\frac{c}{\sqrt{\hbar\omega_{\bot}/m}}=\left(\frac{\tilde{\mu}^2-1}{2 
\tilde{\mu}}\right)
^{\frac{1}{4}} \, ,
 \label{sound}
\end{equation} 
where $\tilde{\mu}=\mu/\hbar\omega_\perp$. This expression provides the speed of
sound for elongated, harmonically trapped condensates with arbitrary values of
the interaction, and gives the exact limits both in the quasi-onedimensional and
Thomas-Fermi regimes.

\begin{figure}[htb]
\center
\includegraphics[width=0.6\linewidth]{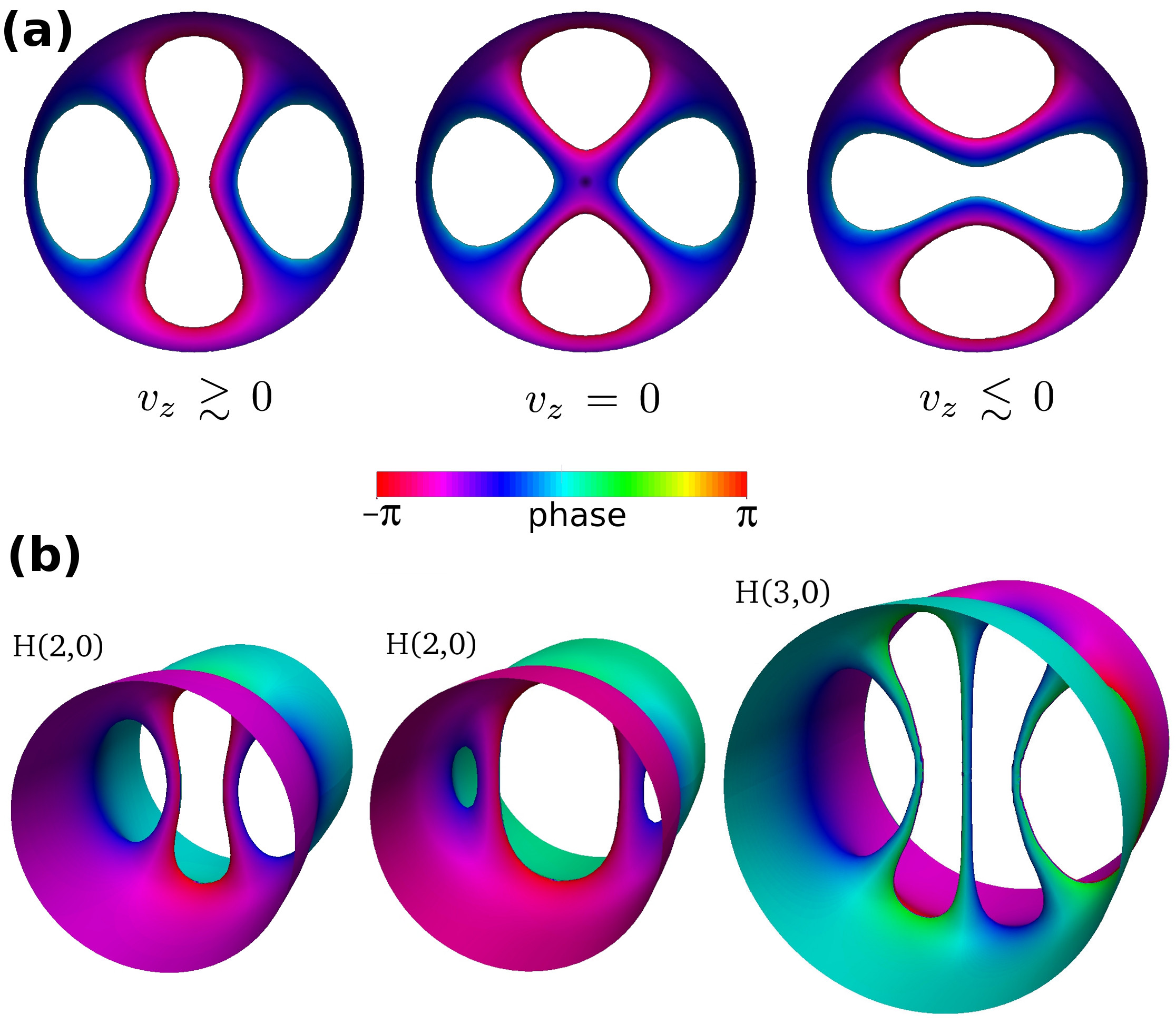}
\caption{ (a) Broken symmetry in a static cross soliton (2SV) with
chemical potential $\mu=5\hbar\omega_\perp$ for small velocity increments
around $v_z=0$, corresponding to points A (at the center),
B (left), and C (right) of Fig. \ref{Fig4}. (b) Hermite-type $(n_x,n_y)$
Chladni solitons: for $\mu=5\hbar\omega_\perp$, H(2,0) solitons have $v_z=0$
(left) and $v_z=0.53\,c$ (right), whereas H(3,0) corresponds to a static 
soliton with $\mu=10\hbar\omega_\perp$. The cross-section of H(2,0) is 
$6.6\,a_\perp$ in width, 
whereas H(2,0) is $9 \,a_\perp$.
}
\label{Hermite}
\end{figure}

Figure \ref{Fig4}b shows the excitation energy $E_s$ as a function of the 
axial velocity $v_z$ for moving solitary waves with chemical potential $\mu=5 
\hbar\omega_\perp$.
Kinks, represented by the solid red line at the top of the figure, 
have the highest excitation energy among the solitary waves, and exist in this 
case only for low velocities, $|v_z|<0.24 \,c$. Below them, at lower 
excitation energies, only solitary waves bifurcating at energies less or equal 
than $\tilde{\mu}=5$ can be found in the graph, as per Eq. (\ref{eq:bif}). As can also be deduced from the 
unstable frequencies of Fig.~\ref{Fig2}, indeed, only vortex rings (blue solid line), 
double solitonic vortex (dashed black) and the single solitonic vortex 
(solid green) are available at $\tilde{\mu}=5$ and appear in Fig.~\ref{Fig4}. At small velocity, and very close in  energy to  the vortex ring and the static cross soliton,
a new type of solution emerges (see the inset of Fig. \ref{Fig4}b). It is composed 
of a couple of almost parallel vortices, indeed a vortex-antivortex pair or vortex
dipole, which is not coming 
directly from a bifurcation of the kink, but from a secondary bifurcation of the cross soliton. Indeed, the solution can be connected to a decay instability of the cross 
soliton produced by the Hermite modes $(n_x=2,n_y=0)$
or $(n_x=0,n_y=2)$.
Figure \ref{Hermite}a shows the density configuration of these states around zero
velocity, which can be cross-checked with their features extracted from the three
panels of Fig.~\ref{Fig4}, where clear differences in the associated phase step 
arise for the three states A, B and C. At higher velocities, the picture is
slighly different. The structure of the kink changes and so do its excitation 
modes. Since the cross soliton is a static state, it can not be found between 
the bifurcations of moving kinks, and is substituted by the mentioned, moving 
Hermite modes. For growing values of the chemical potential, new, higher energy
Hermite modes emerge by equivalent mechanisms. In Fig.~\ref{Hermite}b we show
density isocontours for some of such modes with
$\tilde{\mu}=5, \mbox{and} \, 10$.

It is also interesting to look at the phase jump along the axial $z$-direction 
$\Delta\phi=\phi(z\rightarrow\infty)-\phi(z\rightarrow -\infty)$  created by 
the different solitary waves (Fig.\ \ref{Fig4}a). The phase shift of dark solitons of the one-dimensional nonlinear Schr\"odinger equation (dotted  
line) is shown as a reference. Its phase shows a $\pi$ 
jump in the static configuration, and grows up to $2\pi$ as the velocity 
approaches $+c$, or alternatively reduces to zero as $v_z\rightarrow -c$. 
Similar behaviour, but with different variation rates, is in general followed 
by the phases of 3D solitonic states. However, a particular feature can be 
noted as characteristic of the 3D case. It is the existence of turning points
with vertical phase slopes. Looking at the curves for vortex rings 
(blue lines with triangles), one can notice that there are two separated 
branches ending at respective turning points, and connected by kink states (red 
curve). A more striking manifestation of this phenomenon 
can be observed in the inset of Fig.\ \ref{Fig4}b, on the curve corresponding to 
the vortex-antivortex pair. Between turning points, labeled as B and C in Fig.\ 
\ref{Fig4}c, there exist a set of almost static and degenerate states of this 
type which produce different 
phase jumps. The turning points here indicate a transition between the Hermite-like symmetry of the vortex-antivortex pair and the Laguerre-like vortex cross.
\begin{figure}[htb]
\center
\includegraphics[width=8cm]{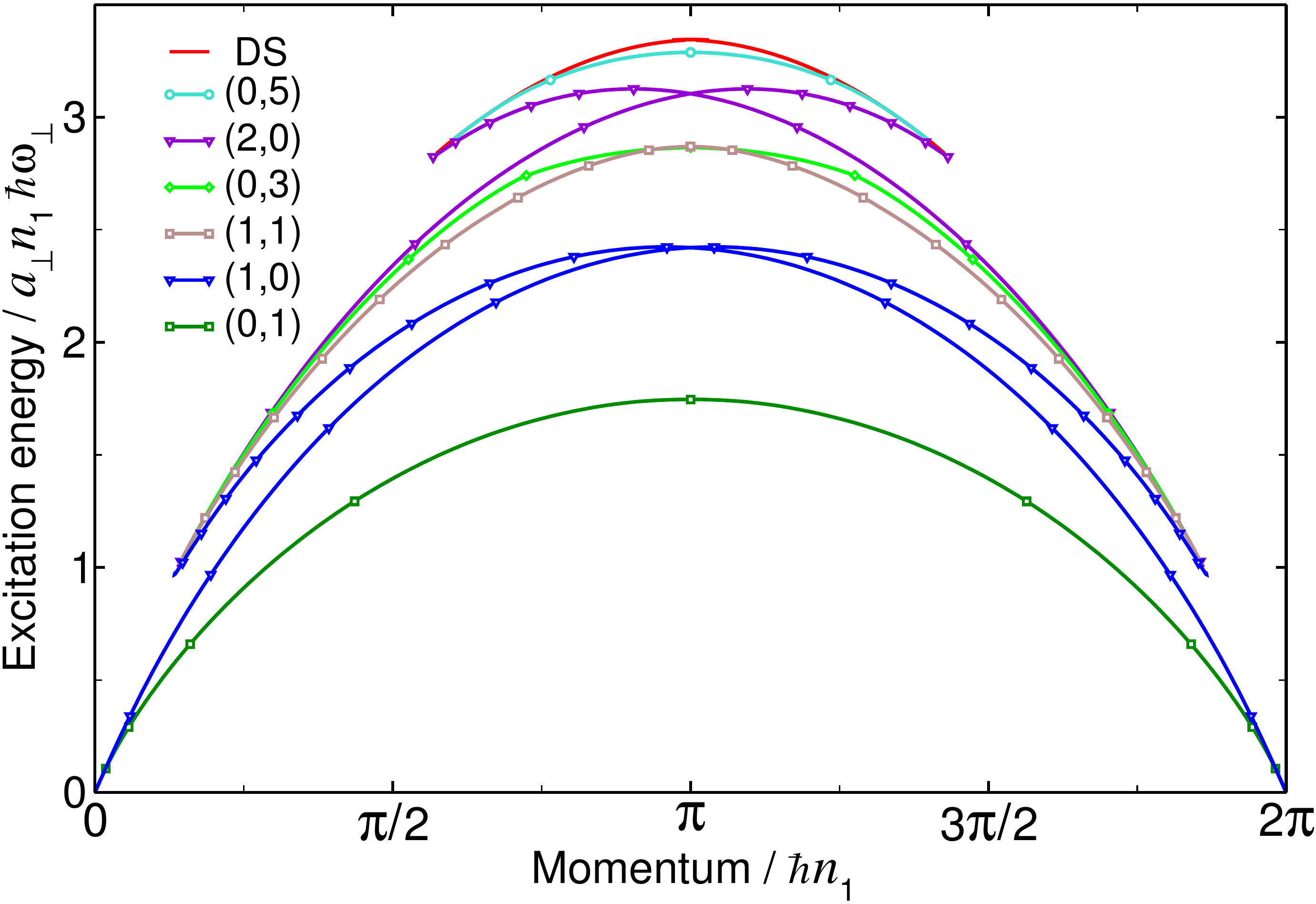}
\caption{Dispersion curves (non-exhaustive diagram) for Chladni
solitons with $\mu=10\hbar\omega_\perp$, some of which correspond to the
isocontour diagrams in Fig.~\ref{Fig3} for static solitons, and in 
Fig.~\ref{moving_solitons} for moving solitons.
}
\label{dispersion_mu10}
\end{figure}

The precedent analysis, in terms of phase and velocity, can now be 
completed by constructing the dispersion relations of Chladni solitons.
To this end, as usual, we define the axial canonical momentum $P_c$ of a 
solitonic state as the conjugate variable of the axial coordinate $z$, that 
fulfils 
\begin{equation}
 v_z=\frac{\partial E_s}{\partial P_{c}} \, .
\end{equation}
Along with the axial physical momentum $p_z=-i\hbar\int d\mathbf{r}(\psi^* 
\partial_z\psi-\psi\partial_z\psi^*)$, carried by the particles traversing 
the plane of the moving soliton, the canonical momentum includes the 
contribution coming from the phase jump $\Delta \phi$ between the axial 
boundaries of the condensate
\begin{equation}
 P_c= p_z -\hbar n_1\Delta \phi \, ,
\end{equation}
where $n_1$ is the axial density of the ground state of the system \cite{Komineas2003a,Scott2011}.
Figure \ref{Fig4}c shows the dispersion relation,
excitation energy versus axial canonical momentum, for solitonic states with
$\mu=5 \hbar\omega_\perp$
moving along the axial coordinate. The turning points that have been described 
on the phase graph, appear here as the vertexes of cusps connecting states with 
the twofold symmetry: vortex rings and vortex-antivortex pairs. It is also 
worth noticing that the lowest excitation energy levels for fixed momentum are
occupied by solitonic vortices, wherever they exist. This last remark accounts 
for the fact that there exist a small regime of soliton speeds, approaching the 
speed of sound, where the only solitonic state is the continuation of the vortex ring family, here an axisymmetric solitary wave very much like a grey soliton with a vortex ring phase singularity outside the Thomas-Fermi density of the trapped BEC, as is 
apparent in Figs. \ref{Fig4}a,b. In this regime, vortex rings are dynamically 
stable states.

\begin{figure}[htb]
\center
\begin{tabular}{@{}cc@{}}
\includegraphics[width=8cm]{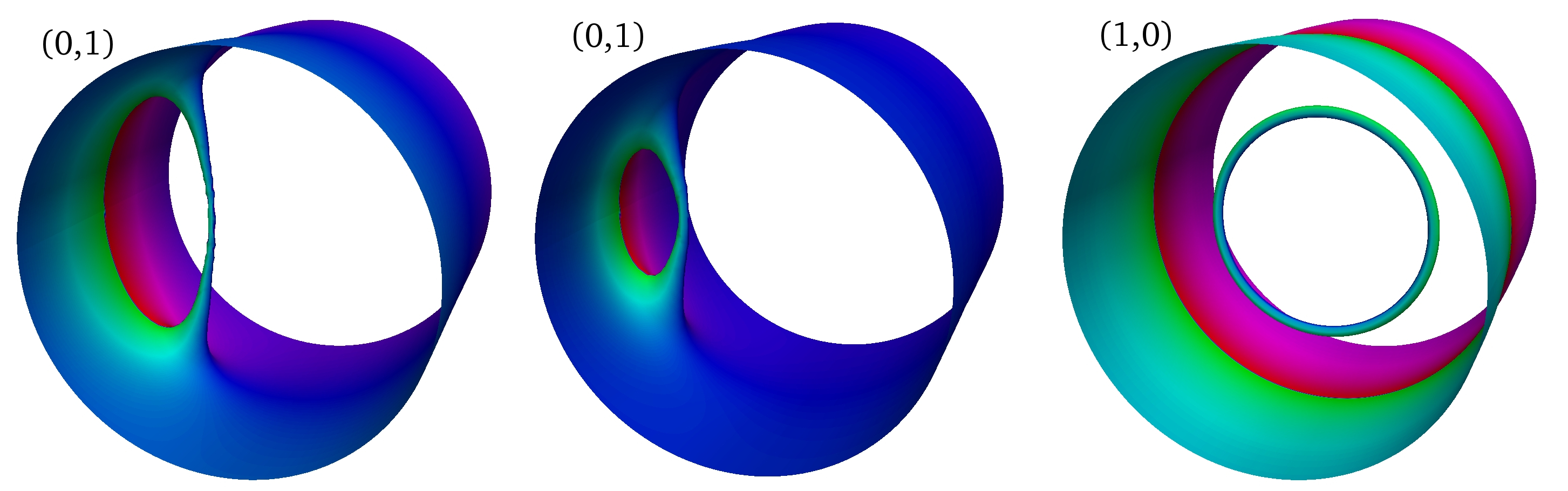}&\hskip-0.3cm
\includegraphics[width=2.6cm]{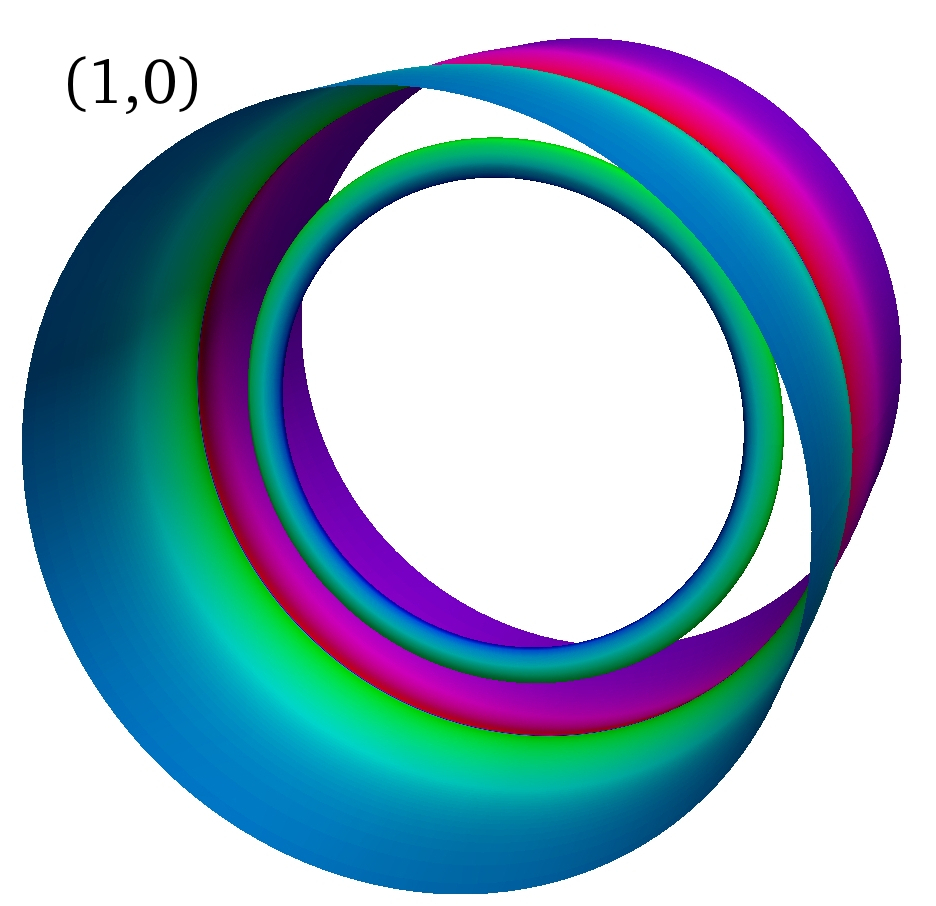}\\ 
\includegraphics[width=8cm]{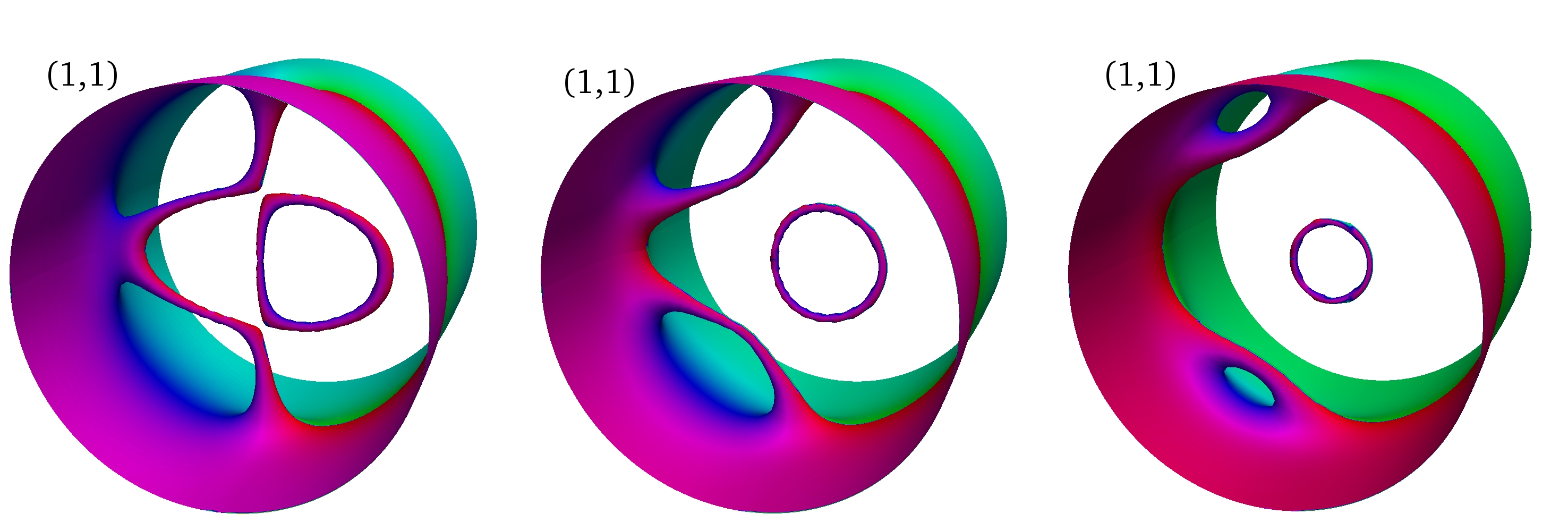}&\hskip-0.3cm
\includegraphics[width=2.6cm]{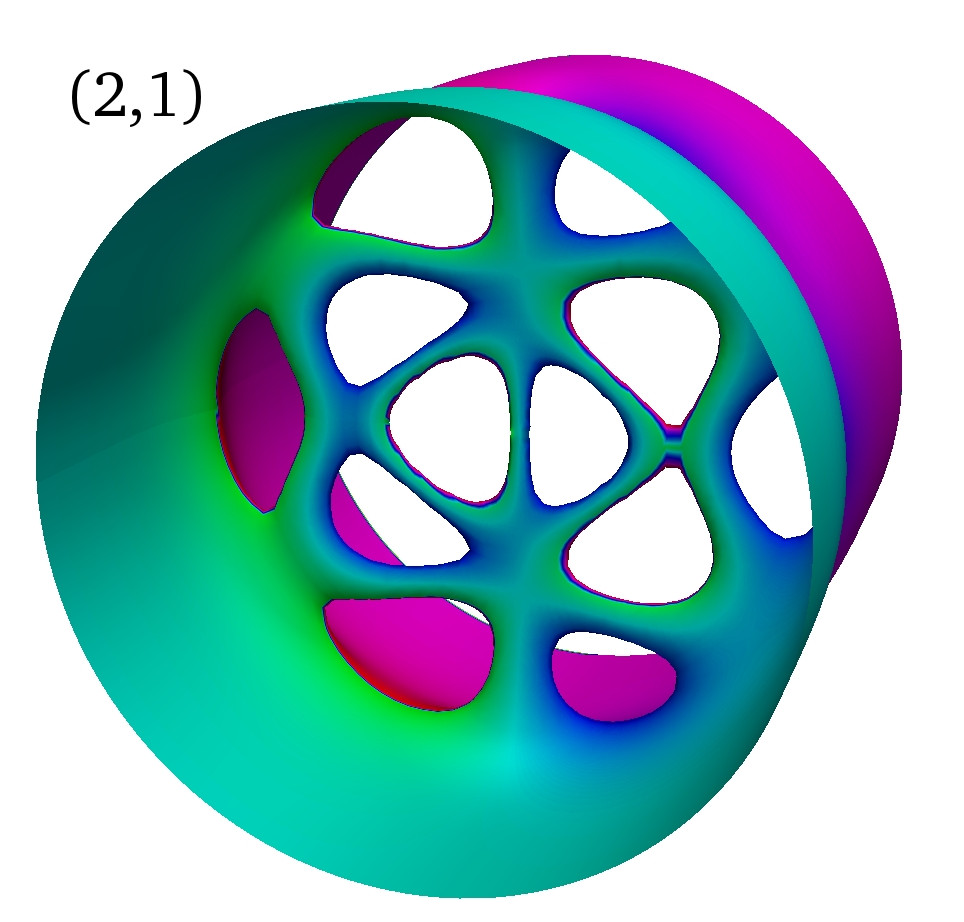}\\
\includegraphics[width=8cm]{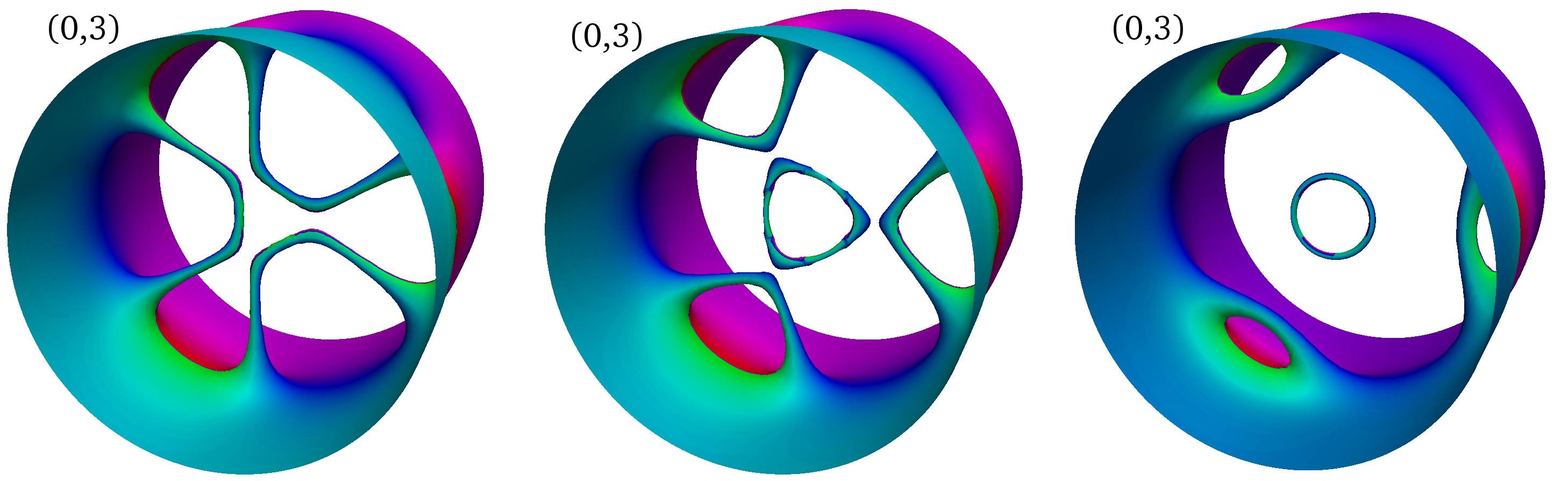}&\hskip-0.2cm
\includegraphics[width=2.7cm]{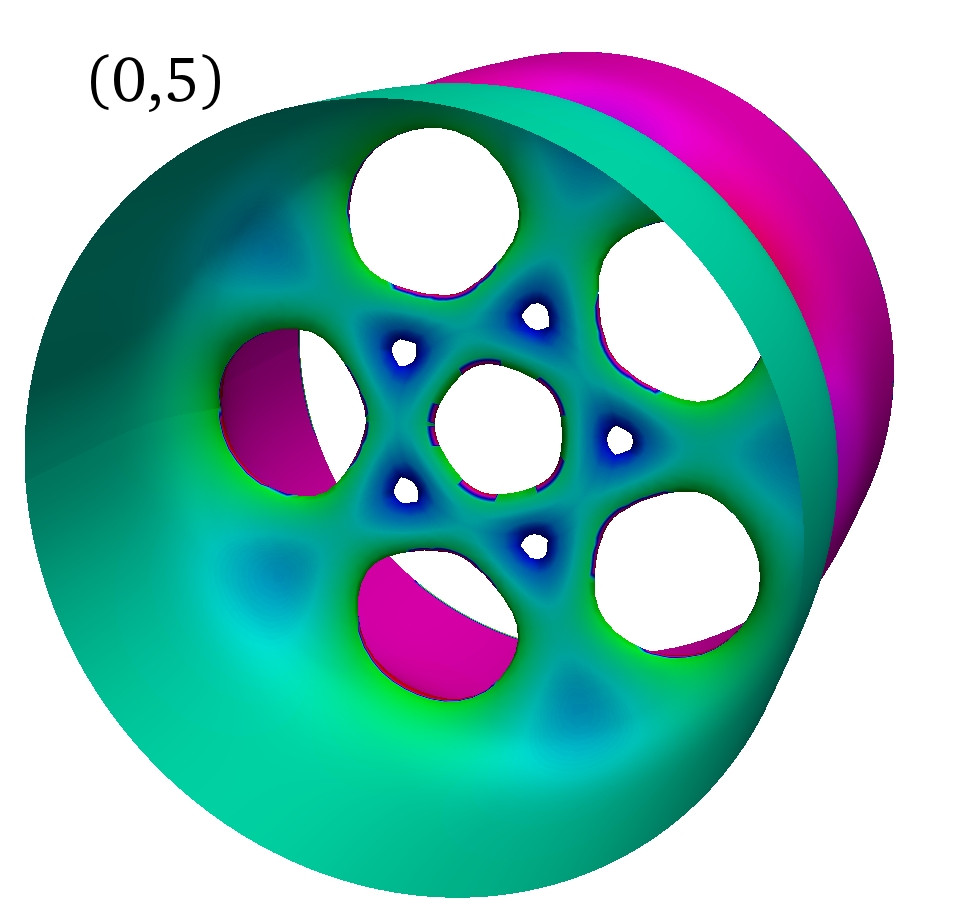}
\end{tabular}
\caption{Steady state configurations of moving Chladni solitons (represented by
density isocontours at 5$\%$ of maximum density) sampling the dispersion curves
in Fig.~\ref{dispersion_mu10} for $\mu=10\hbar\omega_\perp$. The labels indicate
the quantum numbers $(p,l)$ of the associated solitonic family. Several 
isocontours from  the same family correspond to different velocities, with increasing value from left to
right. In this order, the canonical momentum is in units of $\pi\hbar n_1$
for (0,1): 1.97 and 1.99; for (1,0): 1.2 and 1.6; for (1,1): 1.04, 1.3 and 1.7;
for (2,1): 1.06 ; for (0,3): 1.2, 1.3 and 1.6; for (0,5): 1.3. The
cross-sections are $9\,a_\perp$ in width.}
\label{moving_solitons}
\end{figure}

When the chemical potential increases, and the number of bifurcations grows, the 
dispersion diagram of Chladni solitons becomes more complex, because of the 
emergence of new connections between solitary waves sharing symmetries. 
As an instance of this complexity, Fig.~\ref{dispersion_mu10} displays 
some of the curves making the dispersion diagram for $\mu=10 \,\hbar\omega$. 
For this value of the chemical potential, the family of double 
vortex rings [$(p=2,l=0)$,  violet lines] is available, and produces a new couple 
of turning points compared to the case for $\mu=5 \,\hbar\omega$. Following the 
different curves away from their maximum (corresponding to zero 
velocity $v_z=\partial E_s/\partial P_c=0$) the density patterns of Chladni 
solitons can change dramatically from their static configuration. To 
illustrate this point, Figure \ref{moving_solitons} shows the density 
isocontours of some of the moving Chladni solitons that can be found with
$\mu=10 \,\hbar\omega$. It is apparent how their symmetry changes when compared to their static 
counterparts in \ref{Fig3}.

\section{Stability of Chladni solitons}\label{sec:Stability}

\begin{figure}[tb]
\centering
\includegraphics[width=0.75\linewidth]{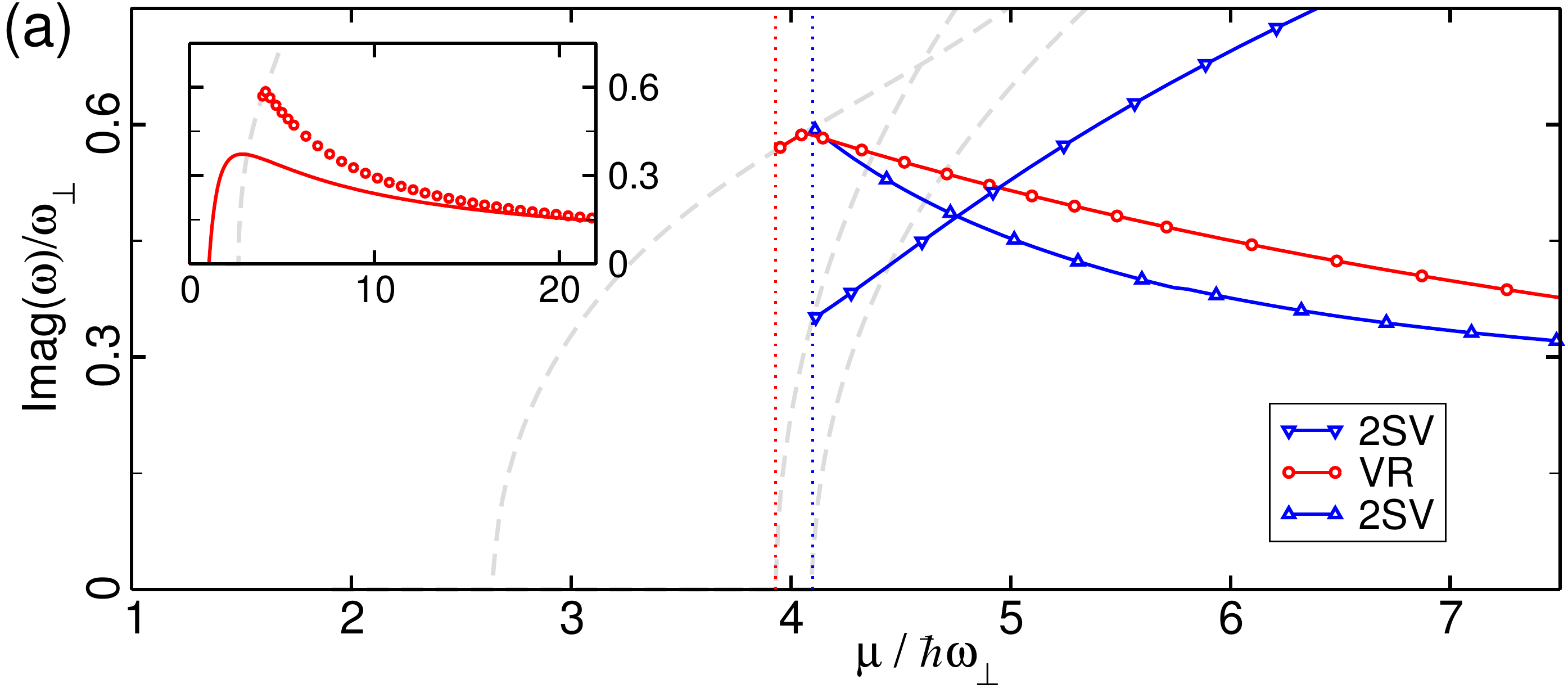}
\vspace{0.1cm}
\includegraphics[width=0.75\linewidth]{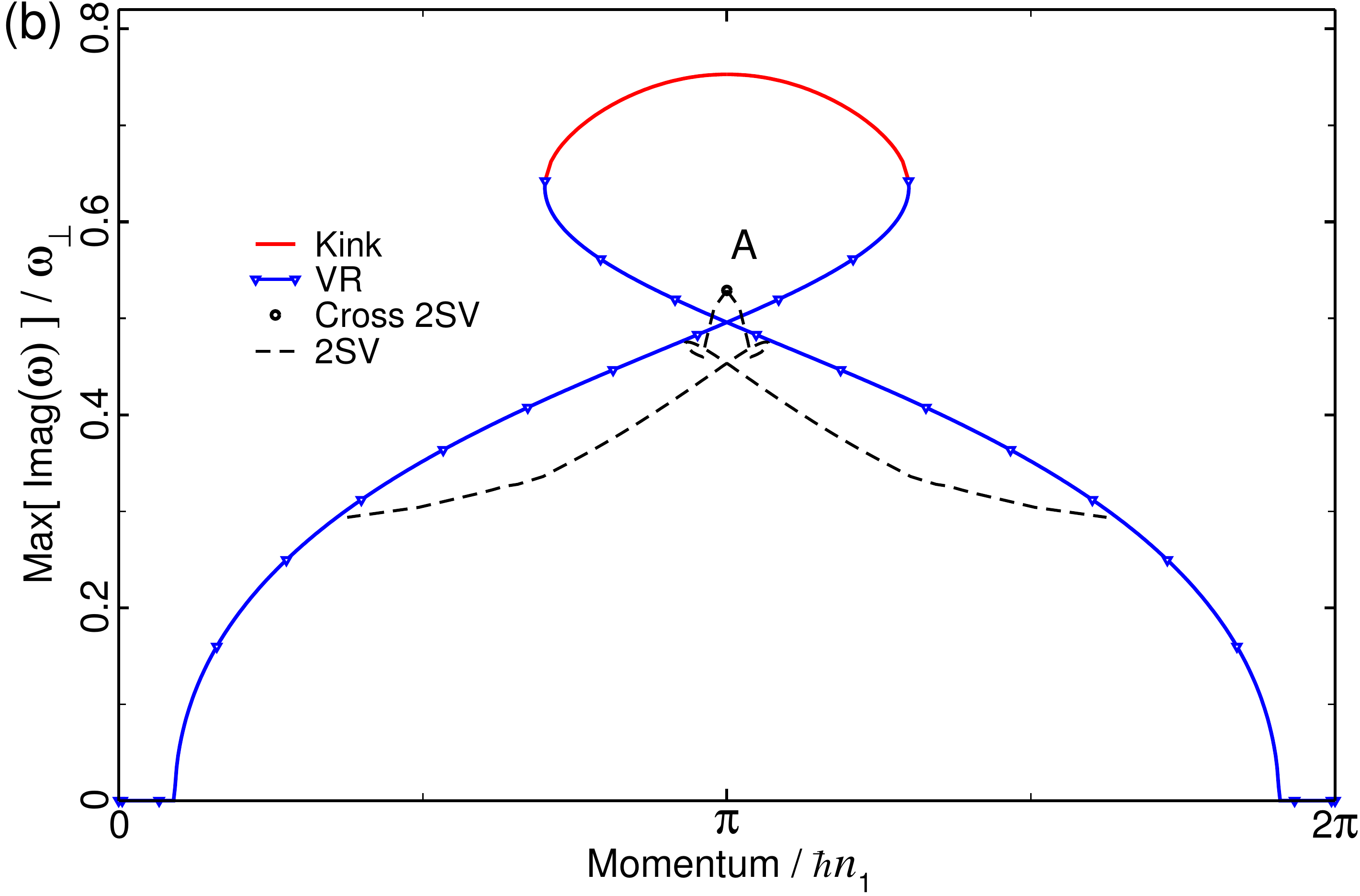}
\caption{(a) Growth rates of unstable modes from numerical solution of Bogoliubov equations \eqref{Bog}, 
for stationary vortex ring (VR) and double solitonic vortex (2SV) states. 
Dotted vertical lines indicate the bifurcation from DS (with unstable 
frequencies represented by dashed grey lines). In the inset, numerical results for vortex rings 
(open symbols) are compared with the analytical prediction \eqref{eq:Horng} (solid line) of Ref.\ 
\cite{Horng2006}. 
(b) Unstable mode growth rates for moving solitary waves with $\mu=5 \hbar 
\omega_\perp$, as a function of the canonical momentum, from a numerical solution 
of the Bogoliubov equations \eqref{Bog}. Point A refers to the cross-vortex 
soliton labeled in Fig.~\ref{Fig4} and represented in the centre 
of Fig.~\ref{Hermite}. }
\label{Fig5}
\end{figure}

It remains to know how robust the Chladni solitons are. To this end, we have 
numerically solved the Bogoliubov equations (\ref{Bog}) for the 
linear excitation modes of the stationary solitonic solutions 
(as those represented in Fig.\ \ref{Fig2}). 
In addition, we have checked the stability of these nonlinear systems against
perturbations by monitoring their evolution in real time through the full 
time dependent Gross-Pitaevskii equation (\ref{3DGPE}).

\subsection{Linear analysis}

As mentioned before, and aside from the stability of vortex rings moving near to 
the speed of sound, our results indicate that the solitonic vortex (SV) branch
is the only one containing dynamically stable states. Solitons corresponding to 
other families are unstable, and decay through the instability channels opened 
by lower energy branches. 
Specifically for the stationary solitary waves,  the solitonic vortex as the lowest energy solitary wave has no channel of instability, since there is no other, lower energy solution bifurcating from it (or from the kink).
However, the second excited state, which is the single vortex ring (VR), does
present one instability channel associated to the bifurcation of solitonic vortices from the kink with lower energy. The next family is that of cross solitons (2SV), and is unstable through
two channels, and so on. This analysis for static states is displayed in 
Fig.\ \ref{Fig5}a. The red curve with open circles corresponds to the unstable
frequencies for vortex rings as a function of the chemical potential, 
and the two blue curves with open triangles indicate the two instability 
channels for the cross soliton. The dotted vertical lines mark the 
bifurcation points for the Chladni solitons considered (VR and 2SV), by
intersecting the instability curves of kinks (gray dashed lines). 
It is worth to mention that, as can be seen in Fig. \ref{Fig5}a, for
intermediate values of the chemical potential, between  
4 and $5 \hbar\omega_\perp$, the cross soliton (2SV, dashed black curve) has smaller values of
unstable frequencies than vortex rings. This fact suggests that cross solitons are good 
candidates for being experimentally realised in elongated BECs. Vortex rings have already
been observed in experiments \cite{Anderson2001,Dutton2001}. In this regard, we have
noticed that in Ref.\ \cite{Becker2013}, where the decay
of dark solitons in anisotropic cigar-shaped condensates was observed in experiments, travelling solitary waves composed of vortex-antivortex pairs (see below)
were clearly identified, and a cross soliton structure was found at the turning 
points of motion.

Additional remarks about vortex ring states are in order. As shown in 
Fig.\ \ref{Fig5}a for the static case, vortex rings are unstable against decay modes with
quantum numbers $(p=0,l=1)$. Our numerical results (open circles in the inset of
Fig.\ \ref{Fig5}a) show that this instability decreases at slow rate with
increasing chemical potential, in agreement with the analytical prediction of
Ref. \cite{Horng2006} (solid red line in the inset):
\begin{equation} \label{eq:Horng}
\omega=-\frac{3\hbar}{\sqrt{2}} 
\frac{\ln[\xi^2(R_\perp^{-2}+\kappa^2/8)]}{2mR_\perp^2} \, ,
\end{equation}
where $\kappa$ is the curvature of the vortex, $\xi$ is the healing length,
and $R_\perp$ is the Thomas-Fermi radius. This expression is valid for vortex rings in
harmonically trapped elongated condensates within the 
Thomas-Fermi regime, and gives nonzero unstable frequencies $\omega$ in the 
limit of very high chemical potential, thus providing an estimation for the
life time of vortex rings. For instance, for 
$\mu=21 \,\hbar \omega_\perp$ both the numerical and analytical methods
predict an unstable frequency $\omega \simeq 0.16 \, \omega_\perp$ corresponding to a life time of about 20 ms for 50-Hz transverse trap.

In the case of moving solitons, the linear stability analysis follows essentially 
the preceding procedure for static states. Figure \ref{Fig5}b, generated for 
$\mu=5 \hbar \omega_\perp$, shows our numerical result for the unstable 
frequencies of moving Chladni solitons as a function of the canonical momentum.
The unstable frequencies decrease rapidly for vortex rings (blue lines with open triangles) of increasing speed (and thus their life time increases), and indeed they become stable past the bifurcation with solitonic vortices close to $P_c=0.2 \,\pi\hbar n_1$, (and $P_c=1.8 \,\pi\hbar n_1$) where the speed approaches the sound speed.
As anticipated, there are also 
no unstable frequencies for solitonic vortices.
As it is the case for the cross soliton, moving vortex dipoles
(black dashed curves) present lower unstable frequencies than vortex rings. This may seem surprising 
considering the cylindrical symmetry of the system, and gives support to their
possible detection in experiments.

\subsection{Real time evolution}

\begin{figure}[tb]
\center
\includegraphics[width=0.85\linewidth]{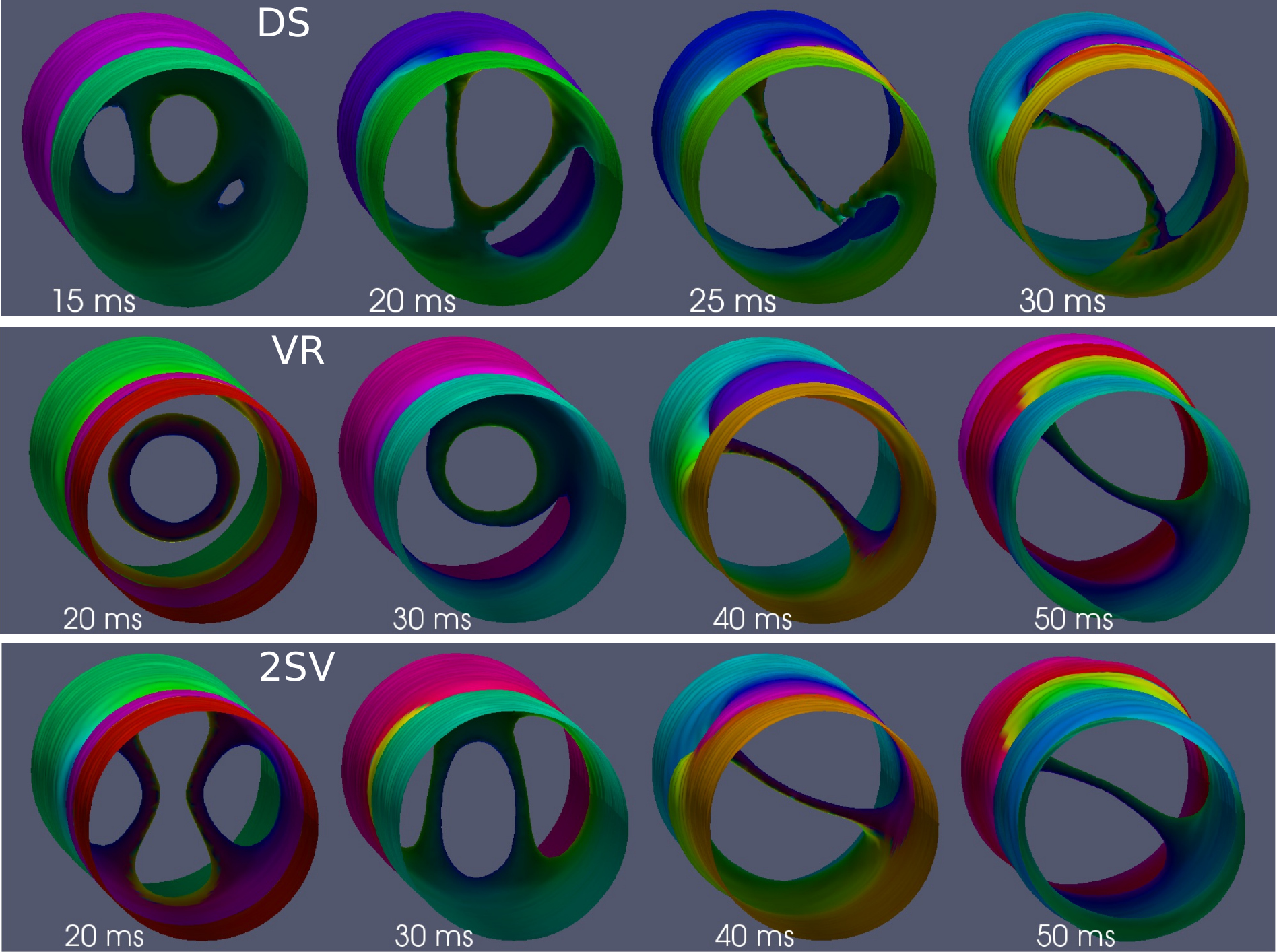}
\caption{ Density isocontours (at 5$\%$ of maximum density) during real time 
evolution showing the decay dynamics of a kink (DS), a vortex ring (VR) and a cross soliton (2SV), with 
$\mu/\hbar \omega_\perp=5$ and $v_z=0$, obtained by solving
numerically the time-dependent Gross-Pitaevskii equation starting from the stationary configuration seeded with a small amount of numerical noise. The typical cross-section
radius is $3.5 a_\perp$.}
\label{Fig6}
\end{figure}
In order to check the predictions given by the linear stability analysis, 
we have also tested the nonlinear stability of Chladni solitons by the
real time evolution of their stationary configurations. For, on the initial states 
$\Psi(\mathbf{r},t=0)$, we have added a random noise perturbation 
$\delta \Psi(\mathbf{r})$, which typically amounts to $2\%$ of the wave
function amplitude. Afterwards, we have allowed these wave functions to evolve in time, 
without dissipation, at constant chemical potential according to Eq.~\eqref{3DGPE}. 
For example, we have followed this procedure for the static Chladni solitons
with $\mu=5 \,\hbar \omega_\perp$, namely dark soliton, vortex ring, cross soliton
and solitonic vortex, which have been previously characterized in Figs.\ \ref{Fig4}
and Figs.\ \ref{Fig5}b. As expected, we have observed the decay of all
solitonic states except the soltionic vortex, which, as 
a result, emerges at the final stage of the time evolution in all cases. Figure \ref{Fig6} 
summarises the decay processes, showing snapshots of the evolution
at intermediate times. In particular it shows complex patterns localised in the plane of the initial stationary state at intermediate times and the emergence of a single solitionic vortex at late times, while some small amplitude radiation moves away from the solitary wave at the speed of sound.

\begin{figure}[tb]
\center
\includegraphics[width=0.85\linewidth]{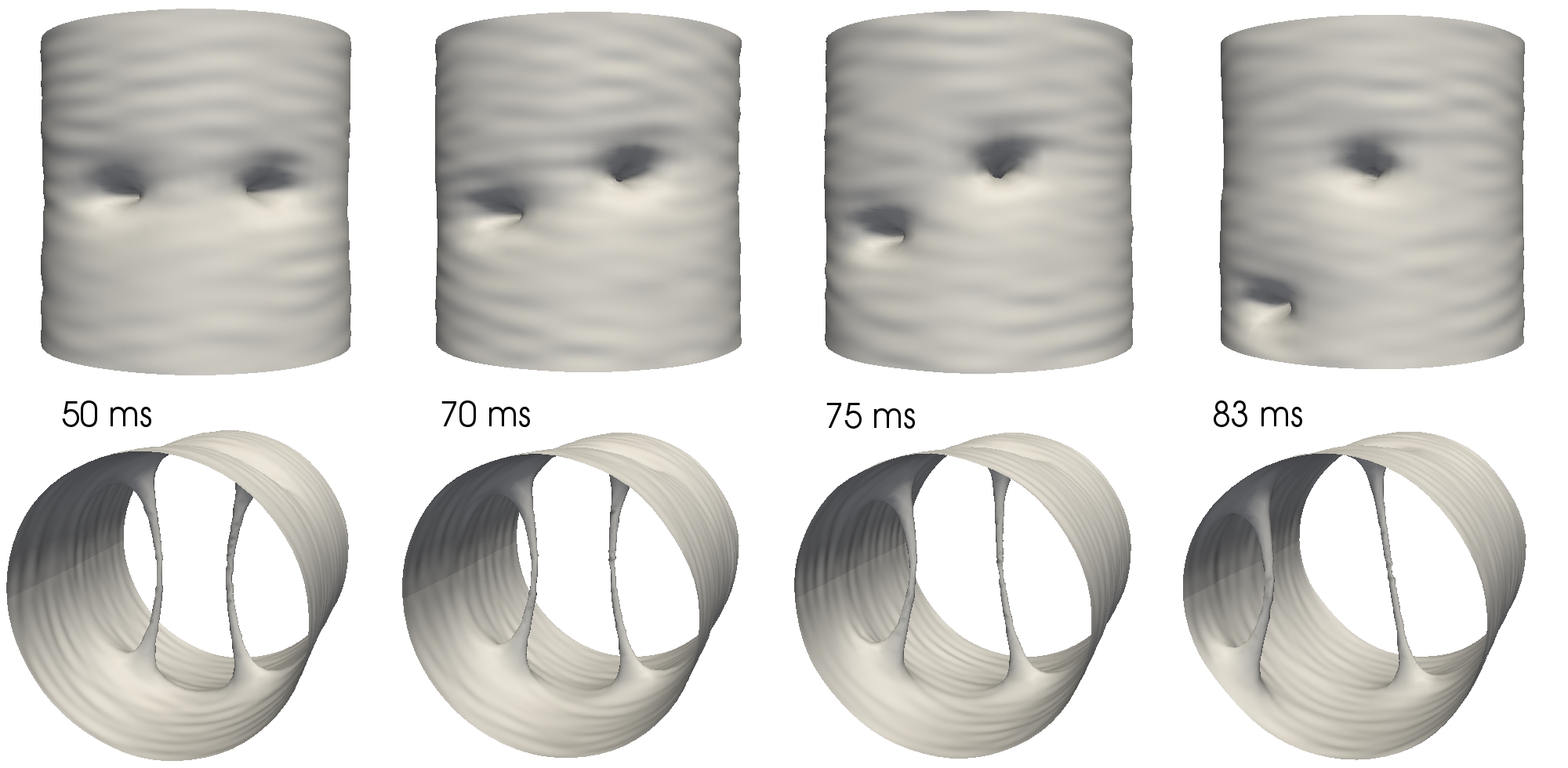}
\vskip0.5cm
\includegraphics[width=0.85\linewidth]{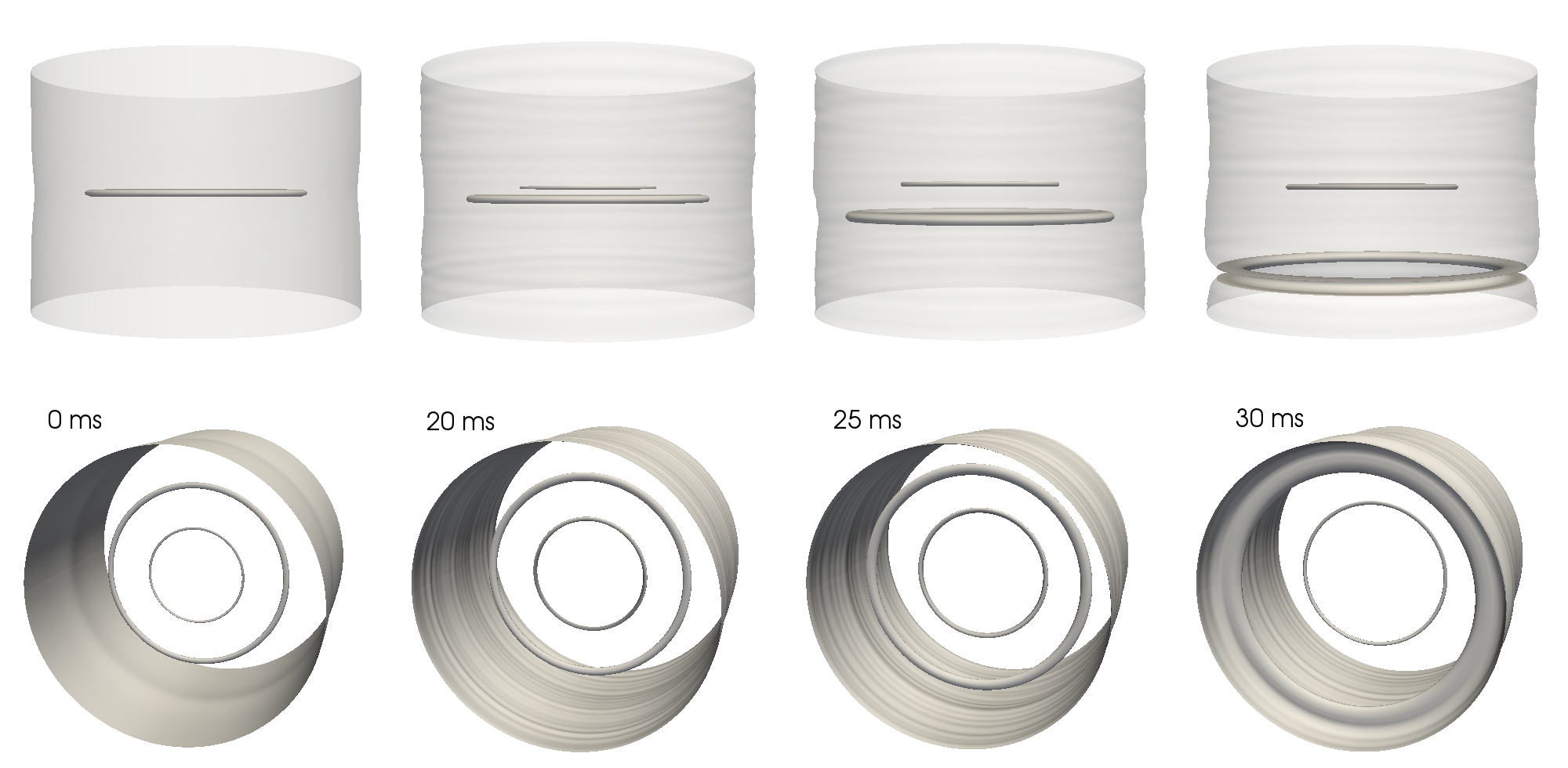}
\caption{ Sling shot events in the real time evolution of a 2SV state (up) with
chemical potential $\mu/\hbar \omega_\perp=8$, and a 2VR state (down) with
$\mu/\hbar \omega_\perp=15$. Density isocontours at 5$\%$ of maximum density
are shown in both cases, with a transparent external surface in the axial view
of the 2VR.}
\label{sling_shots}
\end{figure}
For higher values of the interaction parameter $\tilde{\mu}$, different scenarios can be found in the 
decay of Chladni solitons. In particular more than one solitary wave can appear and move away from the location of the initial unstable soliton.
The interaction energy released from the parent 
state can transform into translational kinetic energy of a descendant solitary 
wave, leaving a simpler structure at the initial position.
This is the case shown at Fig.\ \ref{sling_shots}, which displays 
two types of ``sling-shot'' events, similar to the one recently observed in
experiments and simulations with elongated BECs \cite{Becker2013}. In the upper panels of Fig.\ \ref{sling_shots},
a static vortex-antivortex pair, after a fairly long time scale ($\sim 50$ms),
releases one of the vortices, whose acceleration can be noticed by its 
increasing curvature at later times. As another example, the lower panels of Fig.\ \ref{sling_shots}
follow the evolution of a static double ring that has been perturbed with a
radial noise imprint preserving the cylindrical symmetry. One can observe 
how the external ring escapes from the original transverse plane, and, as in the 
previous case, increases its velocity.

\section{Conclusions}

We have analysed the dynamical properties of static and moving Chladni solitons 
in cylindrically
symmetric Bose-Einstein condensates, within the mean-field regime described
by Gross-Pitaevskii equation. These states, strongly influenced by the geometry
of the trap, emerge from the excitation of standing waves on planar kink states,
and inherit particle-like features characterised by lower excitation energies
and higher inertial masses than the kink. We have calculated numerically 
such quantities, and presented analytical expressions for their evaluation. 
The unstable standing waves producing the decay of the kink have been
object of a detailed analysis, and a formula for the prediction
of the unstable frequencies has been proposed. 

The linear excitations 
and the real time evolution of Chladni solitons have also been addressed. 
Our results suggest that these states, even other than the recently realised
solitonic vortex, could be observed 
in current experiments with ultracold gases, given that their lifetimes
are expected to be of tens of milliseconds. In this regard, several procedures
could be followed. In particular,
in Ref.\ \cite{Mateo2014}, we have proposed a feasible protocol for seeding a particular
Chladni soliton on a planar kink. By means of a dark-bright soliton \cite{Busch2001} in a two component condensate in the
immiscible regime, a proper density and phase pattern could be imprinted on the 
bright soliton of one of the components occupying the kink depletion in the other
component. The subsequent transfer of the selected pattern into the kink component,
through a controlled Raman pulse, could serve the purpose of seeding the decay 
into the corresponding Chladni soliton. 
Other procedures with scalar condensates
relying on an adequate trap geometry, have already been demonstrated.
This is the case in Ref.\ \cite{Becker2013}, where vortex dipoles and the cross soliton has been identified after the decay of kinks in anisotropic 
harmonic traps. Very recently also the $\Phi$ soliton $(p=1,l=1)$ has been identified in what appears to be a seeded decay of a kink in a unitary Fermi gas \cite{Ku}.

\ack
We acknowledge useful discussions with Martin Zwierlein, Shih-Chuan Guo, and P\'eter Jeszenszki.

\section*{References}
\bibliographystyle{unsrt} 
\bibliography{NewRefs.bib,stability_dark_soliton8_njp.bib}

\begin{thebibliography}{10}

\bibitem{Burger1999}
S.~Burger, K.~Bongs, S.~Dettmer, W.~Ertmer, and K.~Sengstock.
\newblock {Dark Solitons in Bose-Einstein Condensates}.
\newblock {\em Phys. Rev. Lett.}, 83(25):5198--5201, December 1999.

\bibitem{Denschlag2000}
J.~Denschlag, J.~E. Simsarian, D.~L. Feder, Charles~W. Clark, L.~A. Collins,
  J.~Cubizolles, L.~Deng, E.~W. Hagley, K.~Helmerson, William~P. Reinhardt,
  S.~L. Rolston, B.~I. Schneider, and William~D. Phillips.
\newblock {Generating Solitons by Phase Engineering of a Bose-Einstein
  Condensate}.
\newblock {\em Science}, 287(5450):97--101, January 2000.

\bibitem{Anderson2001}
B.~P. Anderson, P.~C. Haljan, C.~A. Regal, D.~L. Feder, L.~A. Collins, C.~W.
  Clark, and E.~A. Cornell.
\newblock {Watching Dark Solitons Decay into Vortex Rings in a Bose-Einstein
  Condensate}.
\newblock {\em Phys. Rev. Lett.}, 86(14):2926--2929, April 2001.

\bibitem{ginsberg05}
Naomi Ginsberg, Joachim Brand, and Lene Hau.
\newblock {Observation of Hybrid Soliton Vortex-Ring Structures in
  Bose-Einstein Condensates}.
\newblock {\em Phys. Rev. Lett.}, 94(4):40403, January 2005.

\bibitem{Donadello2014}
Simone Donadello, Simone Serafini, Marek Tylutki, Lev~P Pitaevskii, Franco
  Dalfovo, Giacomo Lamporesi, and Gabriele Ferrari.
\newblock {Observation of Solitonic Vortices in Bose-Einstein Condensates}.
\newblock {\em Phys. Rev. Lett.}, 113:065302, August 2014.

\bibitem{Ku2014}
Mark J.~H. Ku, Wenjie Ji, Biswaroop Mukherjee, Elmer Guardado-Sanchez,
  Lawrence~W Cheuk, Tarik Yefsah, and Martin~W Zwierlein.
\newblock {Motion of a Solitonic Vortex in the BEC-BCS Crossover}.
\newblock {\em Phys. Rev. Lett.}, 113:065301, August 2014.

\bibitem{Mollenauer2006}
Linn~F. Mollenauer and James~P. Gordon.
\newblock {\em {Solitons in Optical Fibers: Fundamentals and Applications}}.
\newblock Academic Press, 2006.

\bibitem{Wright2013}
K.~C Wright, R.~B Blakestad, C.~J Lobb, W.~D Phillips, and G.~K Campbell.
\newblock {Driving Phase Slips in a Superfluid Atom Circuit with a Rotating
  Weak Link}.
\newblock {\em Phys. Rev. Lett.}, 110(2):025302, January 2013.

\bibitem{Eckel2014}
Stephen Eckel, Jeffrey~G. Lee, Fred Jendrzejewski, Noel Murray, Charles~W.
  Clark, Christopher~J. Lobb, William~D. Phillips, Mark Edwards, and
  Gretchen~K. Campbell.
\newblock {Hysteresis in a quantized superfluid ‘atomtronic’ circuit}.
\newblock {\em Nature}, 506(7487):200--203, February 2014.

\bibitem{Mateo2015}
A.~Mu\~{n}oz Mateo, A.~Gallem\'{\i}, M.~Guilleumas, and R.~Mayol.
\newblock {Persistent currents supported by solitary waves in toroidal
  Bose-Einstein condensates}.
\newblock {\em Phys. Rev. A}, 91:063625, 2015.

\bibitem{Frantzeskakis2010}
D.~J. Frantzeskakis.
\newblock {Dark solitons in atomic Bose-Einstein condensates: from theory to
  experiments}.
\newblock {\em J. Phys. A. Math. Gen.}, 43:213001, 2010.

\bibitem{Kanamoto2008}
Rina Kanamoto, Lincoln Carr, and Masahito Ueda.
\newblock {Topological Winding and Unwinding in Metastable Bose-Einstein
  Condensates}.
\newblock {\em Phys. Rev. Lett.}, 100(6):060401, February 2008.

\bibitem{Muryshev1999}
A.~E. Muryshev, H.~B. {van Linden van den Heuvell}, and G.~V. Shlyapnikov.
\newblock {Stability of standing matter waves in a trap}.
\newblock {\em Phys. Rev. A}, 60:R2665--R2668, October 1999.

\bibitem{brand01a}
Joachim Brand and William~P Reinhardt.
\newblock {Generating ring currents, solitons and svortices by stirring a
  Bose-Einstein condensate in a toroidal trap}.
\newblock {\em J. Phys. B At. Mol. Opt. Phys.}, 34:L113--L119, February 2001.

\bibitem{Brand2002}
Joachim Brand and William~P. Reinhardt.
\newblock {Solitonic vortices and the fundamental modes of the “snake
  instability”: Possibility of observation in the gaseous Bose-Einstein
  condensate}.
\newblock {\em Phys. Rev. A}, 65:043612, April 2002.

\bibitem{Mateo2014}
A.~{Mu\~{n}oz Mateo} and J~Brand.
\newblock {Chladni Solitons and the Onset of the Snaking Instability for Dark
  Solitons in Confined Superfluids}.
\newblock {\em Phys. Rev. Lett.}, 113(25):255302, December 2014.

\bibitem{Chladni1787}
Ernst Florens~Friedrich Chladni.
\newblock {\em {Entdeckungen \"{u}ber die Theorie des Klanges}}.
\newblock Weidmanns Erben und Reich, Leipzig, 1787.

\bibitem{Komineas2003}
S.~Komineas and N.~Papanicolaou.
\newblock {Solitons, solitonic vortices, and vortex rings in a confined
  Bose-Einstein condensate}.
\newblock {\em Phys. Rev. A}, 68:043617, October 2003.

\bibitem{Komineas2002}
S.~Komineas and N.~Papanicolaou.
\newblock {Vortex Rings and Lieb Modes in a Cylindrical Bose-Einstein
  Condensate}.
\newblock {\em Phys. Rev. Lett.}, 89:070402, July 2002.

\bibitem{Komineas2003a}
S.~Komineas and N.~Papanicolaou.
\newblock {Nonlinear waves in a cylindrical Bose-Einstein condensate}.
\newblock {\em Phys. Rev. A}, 67:023615, February 2003.

\bibitem{Ku}
Mark J~H Ku, Biswaroop Mukherjee, Tarik Yefsah, and Martin~W Zwierlein.
\newblock {From Planar Solitons to Vortex Rings and Lines: Cascade of Solitonic
  Excitations in a Superfluid Fermi Gas}.
\newblock {\em Arxiv Prepr. arXiv1507.01047}, 2015.

\bibitem{Konotop2004}
Vladimir~V. Konotop and Lev Pitaevskii.
\newblock {Landau Dynamics of a Grey Soliton in a Trapped Condensate}.
\newblock {\em Phys. Rev. Lett.}, 93:240403, December 2004.

\bibitem{Yefsah}
Tarik Yefsah, Ariel~T Sommer, Mark J~H Ku, Lawrence~W. Cheuk, Wenjie Ji,
  Waseem~S Bakr, and Martin~W Zwierlein.
\newblock {Heavy solitons in a fermionic superfluid.}
\newblock {\em Nature}, 499:426--30, July 2013.

\bibitem{Reichl2013}
Matthew~D Reichl and Erich~J Mueller.
\newblock {Vortex ring dynamics in trapped Bose-Einstein condensates}.
\newblock {\em Phys. Rev. A}, 88:053626, 2013.

\bibitem{Bulgac2013}
Aurel Bulgac, Michael~Mcneil Forbes, Michelle~M. Kelley, Kenneth~J Roche, and
  Gabriel Wlaz{\l}owski.
\newblock {Quantized Superfluid Vortex Rings in the Unitary Fermi Gas}.
\newblock {\em Phys. Rev. Lett.}, 112(2):025301, January 2014.

\bibitem{Scott2011}
R.~Scott, F.~Dalfovo, L.~Pitaevskii, and S.~Stringari.
\newblock {Dynamics of Dark Solitons in a Trapped Superfluid Fermi Gas}.
\newblock {\em Phys. Rev. Lett.}, 106:185301, May 2011.

\bibitem{Busch2000}
Th. Busch and J.~R. Anglin.
\newblock {Motion of Dark Solitons in Trapped Bose-Einstein Condensates}.
\newblock {\em Phys. Rev. Lett.}, 84(11):2298--2301, March 2000.

\bibitem{Liao11pr:FermiSolitons}
Renyuan Liao and Joachim Brand.
\newblock {Traveling dark solitons in superfluid Fermi gases}.
\newblock {\em Phys. Rev. A}, 83:041604(R), April 2011.

\bibitem{Becker2008}
Christoph Becker, Simon Stellmer, Parvis Soltan-Panahi, S\"{o}ren D\"{o}rscher,
  Mathis Baumert, Eva-maria Richter, Jochen Kronj\"{a}ger, Kai Bongs, and Klaus
  Sengstock.
\newblock {Oscillations and interactions of dark and dark–bright solitons in
  Bose–Einstein condensates}.
\newblock {\em Nat. Phys.}, 4(6):496--501, May 2008.

\bibitem{Serafini2015}
Simone Serafini, Matteo Barbiero, Michele Debortoli, Simone Donatello, Fabrizio
  Larcher, Franco Dalfovo, Giacomo Lamporesi, and Gabriele Ferrari.
\newblock {Dynamics and interaction of vortex lines in an elongated
  Bose-Einstein condensate}.
\newblock {\em Arxiv Prepr. arXiv1507.01511}, 2015.

\bibitem{Mateo2007}
A~Mu\~{n}oz Mateo and V~Delgado.
\newblock {Ground-state properties of trapped Bose-Einstein condensates:
  Extension of the Thomas-Fermi approximation}.
\newblock {\em Phys. Rev. A}, 75(6):063610, June 2007.

\bibitem{Rosen1932}
N~Rosen and Philip~M Morse.
\newblock {On the vibrations of polyatomic molecules}.
\newblock {\em Phys. Rev.}, 42:210, 1932.

\bibitem{Kamchatnov2008}
A.~Kamchatnov and L.~Pitaevskii.
\newblock {Stabilization of Solitons Generated by a Supersonic Flow of
  Bose-Einstein Condensate Past an Obstacle}.
\newblock {\em Phys. Rev. Lett.}, 100(16):160402, April 2008.

\bibitem{Kuznetsov1988}
EA~Kuznetsov and SK~Turitsyn.
\newblock {Instability and collapse of solitons in media with a defocusing
  nonlinearity}.
\newblock {\em Sov. Phys. JETP}, 67(August 1988):1583--1588, 1988.

\bibitem{Mateo2006}
A.~{Mu\~{n}oz Mateo} and V~Delgado.
\newblock {Extension of the Thomas-Fermi approximation for trapped
  Bose-Einstein condensates with an arbitrary number of atoms}.
\newblock {\em Phys. Rev. A}, 74(6):065602, December 2006.

\bibitem{Horng2006}
T.-L. Horng, S.-C. Gou, and T.-C. Lin.
\newblock {Bending-wave instability of a vortex ring in a trapped Bose-Einstein
  condensate}.
\newblock {\em Phys. Rev. A}, 74(4):041603, October 2006.

\bibitem{Dutton2001}
Z~Dutton, M~Budde, C~Slowe, and L~V Hau.
\newblock {Observation of quantum shock waves created with ultra- compressed
  slow light pulses in a Bose-Einstein condensate.}
\newblock {\em Science}, 293(5530):663--8, July 2001.

\bibitem{Becker2013}
C.~Becker, K.~Sengstock, P.~Schmelcher, P.~G. Kevrekidis, and
  R~Carretero-Gonz\'{a}lez.
\newblock {Inelastic collisions of solitary waves in anisotropic
  Bose–Einstein condensates: sling-shot events and expanding collision
  bubbles}.
\newblock {\em New J. Phys.}, 15(11):113028, November 2013.

\bibitem{Busch2001}
Th. Busch and J.~Anglin.
\newblock {Dark-Bright Solitons in Inhomogeneous Bose-Einstein Condensates}.
\newblock {\em Phys. Rev. Lett.}, 87(1):010401, June 2001.

\end{thebibliography}

\end{document}